%% file: Kernels+PDE_6.tex
\documentclass[12pt]{article}
\usepackage{amsmath}
\usepackage{amssymb}
\usepackage{bbm}
 \usepackage[hmargin=3.54cm,tmargin=3.7cm,bmargin=3.7cm]{geometry}
 \usepackage{color}

\numberwithin{equation}{section}

 \usepackage{graphicx,psfrag}%

\usepackage{graphicx}

\usepackage{pict2e}%

 \usepackage{pdfsync}

\newcommand{\MAT}[1]{\left(\begin{array}{*#1c}}
\newcommand{\mat}{\end{array}\right)}
\newcommand{\qed}{\leavevmode\unskip\nobreak\penalty200\hskip2pt\null
\nobreak\hfill\rule{1.1ex}{1.1ex}
\medbreak }

\newcommand{\Z}{\mathbb{Z}}
\newcommand{\R}{\mathbb{R}}

\newcommand{\rg}{\rightarrow}

\newcommand{\DF}{\Longleftrightarrow}

\newcommand{\AR}{{\cal A}}

\newcommand{\DR}{{\cal D}}

\newcommand{\LR}{{\cal L}}

\newcommand{\PR}{{\cal P}}

\newcommand{\WR}{{\cal W}}

\newcommand{\BC}{{\mathbb C}}

\newcommand{\BQ}{{\mathbb Q}}

\newcommand{\BX}{{\mathbb X}}
\newcommand{\BY}{{\mathbb Y}}
\newcommand{\BZ}{{\mathbb Z}}

\newcommand{\iy}{\infty}
\newcommand{\pl}{\partial}

\newcommand{\Ai}{{\mathrm{Ai}}}

\newcommand{\un}{\mbox{1\hspace{-3.2pt}I}}

\newcommand{\no}{\nonumber}

\newenvironment{proof}{\medskip\noindent{\it Proof:\/} }{\qed}
\newenvironment
        {example}{\medskip\noindent\underline{\it Example:\/} }{\medbreak}

\newcommand{\om}{\omega}
\newcommand{\vp}{\varphi}

\newcommand{\dt}{\delta}

\newcommand{\vr}{\varepsilon}

\newcommand{\BR}{{\mathbb R}}

\newcommand{\lb}{\lambda}

\newcommand{\dis}{\displaystyle}

\newcommand{\rr}{\operatorname{tr}}

\newcommand{\diag}{\operatorname{diag}}

\def\be#1\ee{\begin{equation}#1\end{equation}}
\def\bea#1\eea{\begin{eqnarray}#1\end{eqnarray}}
\def\bean#1\eean{\begin{eqnarray*}#1\end{eqnarray*}}

\newtheorem{definition}{Definition}[section]
\newtheorem{theorem}[definition]{Theorem}
\newtheorem{lemma}[definition]{Lemma}
\newtheorem{corollary}[definition]{Corollary}
\newtheorem{proposition}[definition]{Proposition}

\catcode `!=11

\newdimen\squaresize
\newdimen\thickness
\newdimen\Thickness
\newdimen\ll! \newdimen \uu! \newdimen\dd! \newdimen \rr! \newdimen
\temp!

\def\sq!#1#2#3#4#5{%
\ll!=#1 \uu!=#2 \dd!=#3 \rr!=#4
\setbox0=\hbox{%
 \temp!=\squaresize\advance\temp! by .5\uu!
 \rlap{\kern -.5\ll!
 \vbox{\hrule height \temp! width#1 depth .5\dd!}}%
 \temp!=\squaresize\advance\temp! by -.5\uu!
 \rlap{\raise\temp!
 \vbox{\hrule height #2 width \squaresize}}%
 \rlap{\raise -.5\dd!
 \vbox{\hrule height #3 width \squaresize}}%
 \temp!=\squaresize\advance\temp! by .5\uu!
 \rlap{\kern \squaresize \kern-.5\rr!
 \vbox{\hrule height \temp! width#4 depth .5\dd!}}%
 \rlap{\kern .5\squaresize\raise .5\squaresize
 \vbox to 0pt{\vss\hbox to 0pt{\hss $#5$\hss}\vss}}%
}
 \ht0=0pt \dp0=0pt \box0
}

\def\vsq!#1#2#3#4#5\endvsq!{\vbox to \squaresize{\hrule
width\squaresize height 0pt%
\vss\sq!{#1}{#2}{#3}{#4}{#5}}}

\newdimen \LL! \newdimen \UU! \newdimen \DD! \newdimen \RR!

\def\vvsq!{\futurelet\next\vvvsq!}
\def\vvvsq!{\relax
  \ifx     \next l\LL!=\Thickness \let\continue=\skipnexttoken!
  \else\ifx\next u\UU!=\Thickness \let\continue=\skipnexttoken!
  \else\ifx\next d\DD!=\Thickness \let\continue=\skipnexttoken!
  \else\ifx\next r\RR!=\Thickness \let\continue=\skipnexttoken!
  \else\def\continue{\vsq!\LL!\UU!\DD!\RR!}%
  \fi\fi\fi\fi
  \continue}

\def\skipnexttoken!#1{\vvsq!}

\def\place#1#2#3{\vbox to 0pt{\vss
\rlap{\kern#1\squaresize
  \raise#2\squaresize\hbox{$#3$}}
\vss}}

\catcode `!=12

\long\def\symbolfootnote[#1]#2{\begingroup%
\def\thefootnote{\fnsymbol{footnote}}\footnote[#1]{#2}\endgroup}

\begin{document}

\begin{titlepage}
\begin{flushright}
\end{flushright}
\begin{center}
\begin{Large}
\textbf{Nonlinear PDEs for gap probabilities in random matrices and KP theory.}
\end{Large}

\bigskip
M. Adler\symbolfootnote[2]{Department of Mathematics, Brandeis University, Waltham, MA 02454, USA. adler@brandeis.edu. The support of a National Science Foundation grant \# DMS-07-00782 is gratefully acknowledged}\hspace{0.5cm}
M. Cafasso\symbolfootnote[3]{LAREMA, Universit\'e d'Angers,
2 Bd. Lavoisier 49045 Angers, France. cafasso@math.univ-angers.fr. The hospitality of the Max Planck Institute for Mathematics in Bonn is gratefully acknowledged.}\hspace{.5cm}
P. van Moerbeke\symbolfootnote[4]{D\'epartement	de	Math\'ematiques, Universit\'e de Louvain, 1348 Louvain-la-Neuve, Belgium and Brandeis University, Waltham, MA 02454, USA. pierre.vanmoerbeke@uclouvain.be. The support of a National Science Foundation grant \# DMS-07-00782, a European Science Foundation grant (MISGAM), a Marie Curie Grant (ENIGMA), Nato, FNRS and Francqui Foundation grants is gratefully acknowledged.}
\end{center}
\bigskip
\begin{center}{\bf Abstract}
\end{center}
\qquad Airy and Pearcey-like kernels and generalizations arising in random matrix theory are expressed as double integrals of ratios of exponentials, possibly multiplied with a rational function. In this work it is shown that  such kernels are intimately related to wave functions for polynomial (Gel'fand-Dickey) reductions or rational reductions of the KP-hierarchy; their Fredholm determinant also satisfies linear PDEs (Virasoro constraints), yielding, in a systematic way, non-linear
PDEs for the Fredholm determinant of such kernels. Examples include Fredholm determinants giving the gap probability of some infinite-dimensional diffusions, like the Airy process,
with or without outliers, and the Pearcey process, with or without inliers.
\end{titlepage}
\tableofcontents

\section{Introduction}

The purpose of this paper is to show that, given an interval $E\subset \BR$ (or a disjoint union of them), the Fredholm determinants of kernels of the type:
\be
\begin{aligned}
& K^{(p)}_{t_1,t_2,\ldots,t_{p-1}}(\lb,\lb')= 
 \displaystyle{\frac{1}{(2\pi i)^2}\int_{\Gamma_p^+}du\int_{\Gamma^-_p}dv\frac{{\rm e}^{-V_p(u)+\lb'u}}{{\rm e}^{-V_p(v)
+\lb v}}\left(\frac{u-w}{v-w}\right)^{n}\frac{1}{u-v}},
\end{aligned}
\label{intro1}\ee
where $\Gamma_p^\pm$ are appropriate contours in the complex plane and $V_p(u)$ is a polynomial in $u$ of degree $p+1$, properly parametrized by $t_1,t_2, \ldots, t_{p-1}$, satisfies some non-linear PDEs, which are generated by KP integrable hierarchies. The Fredholm determinants of such kernels describe in some cases the gap probability of certain infinite--dimensional diffusions, like the Airy process, with or without outliers, and the Pearcey one, with or without inliers. This area has been studied by many authors, see for instance \cite{PrSp, JoDetProc, Johansson3, TW-Dyson, ADvM, TWAiryPr, WidomAiry, ACvM2, AvM-Airy-Sine,BertolaCafasso3,BertolaCafasso1} in the case of the Airy process, or \cite{BleKui1, Brezin2, Brezin3, Brezin4, Brezin5, OkoResh-randomproc, TracyWidomPearcey, AOvM, AvM-Pearcey,BertolaCafasso3,BertolaCafasso1} in the case of the Pearcey process.\\

For example, consider $N$ Brownian motions starting and ending at $x = 0$, going from time $t = 0$ to $t = 1$, and conditioned to not intersect. We are interested in the case when $N$ is very large. Then the average mean density, at time $t \in [0,1]$, has its support on the interval $I_t := \left[-\sqrt{2Nt(1-t)},\sqrt{2Nt(1-t)}\right]$, hence forming a set bounded by an ellipse in the $(x,t)$--plane. If we look at any point on the ellipse with an ``Airy microscope'', i.e. with spatial scaling $N^{-1/6}$ and time scaling $N^{-1/3}$ (compatible with Brownian motion), then in the large $N$ limit we find \cite{Johansson3,PrSp} the stationary Airy process $\mathcal A(w)$, whose gap probability is given by the Fredholm determinant
\be
	{\mathbb P}\left(\mathcal A(w) \cap E = \emptyset \right) = \det \left( \un - \chi_EK^{\mathcal A}_{(0,0)}\right)
\label{intro2}\ee
(here and below $\chi_E$ represents the indicator function of $E$)
with
\be
\begin{aligned}
& K^{\mathcal A}_{(n,w)}(\lb,\lb'):= 
 \displaystyle{\frac{1}{(2\pi i)^2}\int_{\Gamma_2^+}du\int_{\Gamma^-_2}dv\frac{{\rm e}^{-\frac{u^3}3+\lb'u}}{{\rm e}^{-\frac{v^3}3
+\lb v}}\left(\frac{u-w}{v-w}\right)^{n}\frac{1}{u-v}},
\end{aligned}
\label{intro3}\ee
a special case ($p=2$) of (\ref{intro1}) and $\Gamma_2^\pm$ are contours passing through the origin and extending to infinity with an angle of $\pi/3$, see also Corollary \ref{contornigiusti} below.\\
It is a consequence of (\ref{intro2}) and Theorem \ref{introtheorem} of this paper, formula (\ref{intro21}), that
$$\BQ = \BQ(E) := \log\mathbb P(\mathcal A(w) \cap E = \emptyset)$$ satisfies the PDE
\be
	\partial^4 \BQ + 6 (\partial^2 \BQ)^2 + (2-4\vr)\partial \BQ = 0
\label{intro4}\ee
where, given a disjoint union of intervals $E := \bigcup^r_{i=1}[a_{2i-1},a_{2i}]\subset\BR$, we define the two differential operators
\be
	\partial := \sum_i \frac{\pl}{\pl a_i}, \quad\quad \vr := \sum_i a_i\frac{\pl}{\pl a_i}.
\label{intro5}\ee

Now consider, in our non--intersecting Brownian motion model described above, the following modification. Instead of requiring all the particles to end, at $t = 1$, at $x = 0$, we allow $n$ outliers to end at $x = {\rm e}^{-\tau_0}\sqrt{N/2}$ and we look with the Airy microscope described above about the point on the ellipse given by
$$
	(x,t_0) = \left(\sqrt{2Nt_0(1-t_0)}, \left(1+{\rm e}^{-\frac{2(\tau_0+w)}{N^{-1/3}}}\right)^{-1} \right).
$$
This yields, in the large $N$ limit, a new process \cite{ADvM}, namely the $n$--Airy process $\mathcal A^{(n)}(w)$. Instead of (\ref{intro2}) we now find, for the gap probability,
\be
	{\mathbb P}\left(\mathcal A^n(w) \cap E = \emptyset \right) = \det \left( \un - \chi_E K^{\mathcal A}_{(n,w)}\right)
\label{intro6}\ee
and using Theorem \ref{introtheorem}, (\ref{intro25}) with $p = 2$, we find that
$$\BQ = \BQ(E) := \log\mathbb P(\mathcal A^{n}(w) \cap E = \emptyset)$$ satisfies the PDE
\be
	\partial^4 \BQ + 6 (\partial^2 \BQ)^2 + (2-4(\vr-w\partial_w))\partial \BQ + 3\partial_w^2 \BQ = 0.
\label{intro7}\ee
Both the equations (\ref{intro4}) and (\ref{intro7}) are new (here and in the paper we denote, given an arbitrary variable $w$, $\partial_w := \frac{\partial}{\partial w}$). In \cite{ADvM} another much more complicated but nevertheless useful PDE for $\BQ$ was found.\\

Another beautiful example, again in the framework of $N$ non--intersecting Brownian motions, involves all particles starting at $t = 0$ at $x = 0$, with half of them ending at $-\sqrt{N/2}$, the other half at $\sqrt{N/2}$. Now for a time $t\in[0,1/2]$ the average mean density is supported on one interval that, for $t> 1/2$, separates into two. In the $(x,t)$--plane the edge of the support is heart--shaped, with a cusp at $(x,t) = (0,1/2)$ and the heart is anchored 	at the points $(0,0),\;(\pm\sqrt{N/2},1)$. If we now look about the cusp at $(0,1/2)$ with a ``Pearcey microscope'', i.e. with spatial scaling $N^{-1/4}$ and time scaling $N^{-1/2}$ (compatible with Brownian motions), then in the large $N$ limit we find \cite{TracyWidomPearcey} the Pearcey process $\mathcal P(\tau)$ with gap probability given again by a Fredholm determinant (here $E$ must be compact)
 
\be
	{\mathbb P}\left(\mathcal P(\tau) \cap E = \emptyset \right) = \det \left( \un - \chi_E K^{\mathcal P}_{(0,0)}\right)
\label{intro8}\ee
with
\be
\begin{aligned}
& K^{\mathcal P}_{(n,w)}(\lb,\lb'):= 
 \displaystyle{\frac{1}{(2\pi i)^2}\int_{\Gamma_3^+}du\int_{\Gamma^-_3}dv\frac{{\rm e}^{-\frac{u^4}4+\tau\frac{u^2}2 - \lb'u}}{{\rm e}^{-\frac{v^4}4 + \tau\frac{v^2}2 - \lb v}}\left(\frac{u+w}{v+w}\right)^{n}\frac{1}{u-v}},
\end{aligned}
\label{intro3}\ee
another special case ($p=3$) of (\ref{intro1}) (see Corollary \ref{contornigiusti} for a description of the contours). In this case we get for
 
$$\BQ = \BQ(E) := \log\mathbb P(\mathcal P(\tau) \cap E = \emptyset)$$
 the following set of equations, obtained from (\ref{intro8}) and Theorem \ref{introtheorem} upon specializing equations (\ref{intro21})--(\ref{intro24}) for $p = 3$ and upon setting $\tau = 2t_2$:
 \be
 	\pl^4\BQ +  6 (\pl^2 \BQ)^2 + 12\pl_\tau^2\BQ - 4\tau\pl^2\BQ = 0
 \label{intro10}\ee
\be
	2\pl_\tau\pl^3\BQ + 12 (\pl_\tau\pl Q)(\pl^2\BQ - \tau/3) + (1 -3\vr + 2\tau\pl_\tau)\pl\BQ = 0
\label{intro11}\ee
\be
	8\pl_\tau^3\BQ + 4\left\{ \pl_\tau\pl\BQ,\pl^2\BQ\right\}_\partial + (\vr -2\tau\pl_\tau - 2)\pl^2\BQ = 0
\label{intro12}\ee
\be
	\pl^4 U + 12 \pl_\tau^2 U + 6 \pl^2 U^2 = 0,\quad U:=\pl^2\BQ - \frac{\tau}2.
\label{intro13}\ee
The last one is just an other form (Boussinesq form) of (\ref{intro10}). Equations (\ref{intro10}) and (\ref{intro11}) are new, while (\ref{intro12}) was first derived in \cite{AOvM}. In the above and throughout this paper, we denote $\left\{f,g \right\}_\pl := f(\pl g) - (\pl f)g$.\\

In the previous Pearcey example, if we allow $n$ inliers to end, for $t = 1$, at $x := -w\left(\frac{N}2\right)^{1/4}$, that would lead, in the large $N$ limit and near the cusp, to a different process, namely (see \cite{ADvM}) the $n$--Pearcey process $\mathcal P^{(n)}(\tau)$. The gap probability of this process (again $E$ is compact) is given by

\be
	{\mathbb P}\left(\mathcal P^{(n)}(\tau) \cap E = \emptyset \right) = \det \left( \un - K^{\mathcal P}_{(n,w)\chi_E}\right),
\label{intro14}\ee
leading by Theorem \ref{introtheorem}, exactly as for Pearcey, to the equations (\ref{intro26})--(\ref{intro29}) for 
$$\BQ = \BQ(E) := \log\mathbb P(\mathcal P^{(n)}(\tau) \cap E = \emptyset),$$
upon setting $\tau = 2 t_2$. All of these four PDEs are new, although the third one was conjectured in \cite{ADvMVa}.\\
It is also worthwhile to mention that the PDEs that have been derived can be used to obtain asymptotic results. For example, as a consequence of (\ref{intro12}), one can show in the style of \cite{ACvM2} the following approximation of the Pearcey process by the Airy process, by looking far out along the Pearcey cusp to yield, for large $\tau$:
\be
	\mathbb P\left(\frac{\mathcal P(\tau) - \frac{2}{27}(3\tau)^{\frac{3}{2}}}{(3\tau)^{\frac{1}{6}}} \, \cap \, (-E) = \emptyset \right) 
	= \mathbb P\left(\mathcal A(0)  \cap \, (-E) = \emptyset \right)\left(1 + \mathcal O(\tau^{-\frac{4}{3}})\right),
\label{intro15}\ee
where as usual $E$ is a finite union of compact intervals (see also \cite{BertolaCafasso1} for a similar result obtained using the non--linear steepest descent method for Riemann--Hilbert problems).\\
 
In order to state the main theorem of this paper with precision, we need the following preliminaries.\\
 Given the root of unity $\om ={\rm e}^{\frac{\pi i}{p+1}}$, consider any subset of rays $\Gamma_p^\pm$ taken from configurations
\be
\Gamma_p^+ \subset \left\{{\cal C}( \om^{2j}),~~~~j\leq \Big[\frac{p+1}{2}\Big]\right\},~~~\Gamma_p^- \subset \left\{{\cal C}( \om^{2j+1}),~~j\leq \Big[\frac{p}{2}\Big]\right\},
\label{In41}\ee
consisting of oriented counter-clockwise contours ${\cal C}(\omega^\ell)=\BR_+\cdot\omega^\ell+\BR_+\cdot\overline{\omega}^\ell$. 
 Given an integer $n \geq 0$, consider the kernel in (\ref{intro1})
\be
\begin{aligned}
& K^{(p)}_{t_1,\ldots,t_{p-1}}(\lb,\lb')= 
 \displaystyle{\frac{1}{(2\pi i)^2}\int_{\Gamma_p^+}du\int_{\Gamma^-_p}dv\frac{{\rm e}^{-V_p(u)+\lb'u}}{{\rm e}^{-V_p(v)
+\lb v}}\left(\frac{u-w}{v-w}\right)^{n}\frac{1}{u-v}},
\end{aligned}
\label{In48}\ee
where $V_p(u)$ is a polynomial
\be\begin{aligned}
V_p(u)&=\frac{u^{p+1}}{p+1}+\sum_0^{p-2}\theta_i\frac{u^{i+1}}{i+1},\quad
\theta_i :=\theta_i(t_1,...,t_{p-1}) \\
&= \frac{u^{p+1}}{p+1}  -t_{p-1} u^{p-1} -t_{p-2} u^{p-2} -(t_{p-3}+\ldots) u^{p-3}-\ldots
-(t_1+\ldots)u
\end{aligned}
\label{In49}\ee
with polynomial coefficients $\theta_i$ in $t$ implicitly given in terms of $t_1,...,t_{p-1}$, by solving the equation $w=V_p'(u)$ for $u$ in terms of a series in large $w$, as in (i) and identifying with another series, as in (ii); thus %
\bea
u&\stackrel{(i)}{=}&w^{\frac1p}-\frac{1}{p}\theta_{p-2}w^{-\frac1p}-\frac{1}{p}\theta_{p-3}w^{-\frac2p}-
\frac{1}{p}\left(-\frac{p-3}{2p}\theta^2_{p-2}+\theta_{p-4}
\right)w^{-\frac3p}\nonumber\\
& &\hspace*{5cm}+\dots +O(w^{-1-\frac1p})\\
&\stackrel{(ii)}{=}&w^{\frac1p}+\frac{1}{p}\sum_1^{ p-1}(p-j)t_{p-j}w^{-\frac{j}{p}}+O(w^{-1-1/p}).\nonumber
\eea

The Hirota bilinear equations for the KP-flow imply two strings of partial differential equations, characterizing the KP $\tau$-function, namely (see Lemma \ref{lemma6})\footnote{The Hirota symbol of two functions $f$ and $g$, associated with any polynomial $p({\mathbf t})$, is given by
$$p(\pl_1,\pl_2,...)f\circ g=p\left(\frac{\pl}{\pl y_1},\frac{\pl}{\pl y_2},\ldots \right)f(t_1+y_1,t_2+y_2,\ldots)g(t_1-y_1,t_2-y_2,\ldots)\big\vert_{y_i=0},$$ while the elementary Schur polynomials are defined by the relation $${\rm e}^{\sum_{i=1}^\infty t_iz^i} = \sum_{\ell = 0}^\infty z^\ell p_\ell(\mathbf t).$$}
\be
\BY_{\ell}:\Bigl(p_{\ell +1}(\pl_{\mathbf t})-\frac{1}{2}\pl_1\pl_{\ell}\Bigr)\tau\circ\tau=0,  ~~~\BY_{1,\ell -1}:\Bigl(\pl_1\pl_{\ell}-\frac{1}{2}\pl_2\pl_{\ell -1}-\pl_1p_{\ell}(\pl_{\mathbf t})\Bigr)\tau\circ\tau=0,
\nonumber\ee

We further remind the reader of the definition (\ref{intro5}) and the previously stated convention that $\partial_i := \frac{\partial}{\partial t_i}$ and $\partial_w = \frac{\partial}{\partial w}$. Now we can finally state our main result.

\begin{theorem}\label{introtheorem}
Each of the Hirota equations
\be
\BY_3,\ldots,~\BY_{p+1},~\BY_{1,4},\ldots,~\BY_{1,p}~~\mbox{and}~~2(p+2)\BY_{p+2}+(p+1)\BY_{1,p+1}, 
\nonumber
\ee
gives rise to a non-linear PDE for the Fredholm determinant of the kernel (\ref{intro1})
\be
\BQ=\BQ_p(t_2,...,t_{p-1};E):=\log\det(I-K^{(p)}_{t_1,\ldots,t_{p-1}}\raisebox{1mm}{$\chi$}{}_{E} )\big\vert_{t_1= 0}.
\label{intro20}\ee
These PDEs only involve the differentials $\pl$ and $\vr$ with regard to the boundary points of $E$ and the $t$-partials $\pl_2,\pl_3,...,\pl_{p-1}$. Some noteworthy examples are:\newpage
\noindent\boxed{\bf Case\;1\, (n=0):}\\

\noindent $\bullet$
{For all $p\geq 2$, $\BQ$ satisfies\\
\underline{the $\BY_3$-equation}: }
\be
\begin{aligned}
\pl^4\BQ+&6(\pl^2\BQ)^2 +\dt_{2,p}(2-4\vr)\pl\BQ
\\
&+(1-\dt_{2,p})\left(3\pl^2_2\BQ-4\left[3 \frac{ p-1}{p} t_{p-1}\pl+(1-\dt_{3,p})\pl_3\right]\pl\BQ\right)=0
\end{aligned} \label{intro21}\ee

\noindent$\bullet$ For all $p\geq 3$, $\BQ$ satisfies the equation (\ref{intro21}) and \\
\underline{the $\BY_4$-equation}:

\be
\begin{aligned}\pl_2\pl^3 \BQ+6(\pl_2\pl\BQ)\left(\pl^2\BQ-\mbox{$\dis\frac{1}{p}$}(p-1)t_{p-1}\right)
+\dt_{3,p}\Bigl((1-3\vr)\pl\BQ+2t_2\pl\pl_2\BQ\Bigr)
\\
+(1-\dt_{3,p})\left(2\pl_2\pl_3\BQ-3\left[\mbox{$\dis\frac{4}{p}$}(p-2)t_{p-2}\pl+(1-\dt_{4,p})\pl_4\right]\pl\BQ\right)=0
\end{aligned}
\label{intro22} \ee
\underline{the $\pl_2\BY_3-\pl\BY_4$-equation}
\be\begin{aligned}
&\pl_2^3\BQ +2\left\{\pl_2\pl\BQ,\pl^2\BQ\right\}_{\pl}-2(1-\dt_{3,p})\pl_2\pl_3\pl\BQ 
+\dt_{3,p}(\varepsilon-2t_2\pl_2-2)\pl^2\BQ \\
&-\frac2p 
(1-\dt_{3,p})\left(( {p-1} )t_{p-1}\pl_2-2( {p-2} )t_{p-2}\pl-\frac p2 (1-\dt_{4,p})\pl\right) \pl^2\BQ=0
\end{aligned}\label{intro23}
\ee
and \underline{the $ \pl^2\BY_3$-equation} (the Boussinesq form of the $\BY_3$-equation)
\be
 \pl^4 U+3\pl_2^2 U+6\pl^2 U^2-4(1-\dt_{3,p})\pl_3\pl U=0,\;\textrm{where}\; U\!:=\!\pl^2\BQ-\dis\frac{p-1}pt_{p-1}\label{intro24}.
\ee
 
\noindent\boxed{\bf Case\; 2\,(n>0):} $\BQ$ satisfies\\

\noindent\underline{For $p = 2$:}
\be
\pl^4\BQ+6(\pl^2\BQ)^2+(2-4(\vr-w\pl_w))\pl\BQ+
 3\pl^2_w\BQ=0
 \label{intro25}\ee
 \underline{For $p = 3$:}
\be
	\pl^4\BQ+6(\pl^2\BQ)^2-8t_2\pl^2\BQ+3\pl^2_2\BQ 
 	+4\pl_w\pl\BQ=0,
 \label{intro26}\ee
\be
(\pl_2\pl^2-2t_2\pl_2-3(\vr-w\pl_w)+1)\pl\BQ+6(\pl^2\BQ)(\pl_2\pl\BQ) -2\pl_2\pl_w\BQ=0,
\label{intro27}\ee
\be
 	(\vr-w\pl_w -2t_2\pl_2-2)\pl^2\BQ+\pl_2^3\BQ+2\{\pl_2\pl\BQ,\pl^2\BQ\}_{\pl}
	+2\pl_2 \pl_w \pl \BQ=0,
\label{intro28}\ee
\be
\mathrm{and}\quad\pl^4 U+3\pl^2_2U+6\pl^2 U^2+4\pl_w\pl   U=0,\qquad U=\pl^2\BQ-\mbox{$\frac{2}{3} $}t_2. 
\label{intro29}
\ee
Notice that for $n=0$, $\BQ$ also satisfies these equations, but without the terms containing $\pl_w$ (as it does not depend on $w$).
\end{theorem}

Theorem \ref{introtheorem} is new, and it has as consequences the PDEs (\ref{intro4}), (\ref{intro7}), (\ref{intro10})--(\ref{intro13}), and (\ref{intro26})--(\ref{intro29}), for the Airy and Pearcey cases, with or without inliers and outliers, as previously discussed. However, at heart, the theorem is a method for yielding PDEs for Fredholm determinants (\ref{intro20}), with kernels of the form (\ref{In48}), some of which were derived explicitly for the benefit of the reader; although clearly our motivation is random matrix theory. Indeed it would be interesting to relate all the kernels in (\ref{In48}), also for $p\geq 4$, to the theory of random matrices (see open questions below). Other methods for computing PDEs for gap probabilities can be found in \cite{ADvM,ADvMVa,AOvM,AvM-Airy-Sine,AvM-Pearcey,TWAiryPr,TW-Dyson,BorodinDeift,ItsHarnad,BertolaCafasso1,BertolaCafasso3,Ru2,Ru3}.\\

The main tool of this paper is the natural integrable deformations of the kernel (\ref{intro1}), which enables us to represent it essentially as an $x$--integral of the product of a KP--hierarchy wave function and its adjoint, and hence in terms of KP tau functions, as elaborated in Proposition \ref{prop4}. Consequently, we can use the the tools of Sato's (formal) KP theory of Grassmannians, which are provided for the reader in Section 2, along with applications of these tools, which we shall need in the next section. In Section 3, we show, using a proposition originally appearing in \cite{ASvM}, how to use the fact that the ``integrable deformation'' of the Fredholm determinant (\ref{intro20}) has an expression as a ratio of two tau functions as in (\ref{44}), (in essence a continuous soliton formula), to derive the so--called ``Virasoro relations'' for the Fredholm determinant itself. This leads in Section 4 to a derivation of the PDEs of Theorem \ref{introtheorem}, or more precisely to a method to derive PDEs. In the appendix we apply the method to derive the first example in Theorem \ref{introtheorem}, leaving the others to the interested reader. There is a large number of formulas in this paper, first going from Sato's theory to its applications, and then in assembling the formulas necessary to derive any PDE inherent in Theorem \ref{introtheorem}. Removing some of the formulas would make it impossible to check the results of the paper and moreover to reproduce these PDEs.\\
We end this introduction with some open questions:
\begin{enumerate}
	\item In this paper we derived PDEs just for scalar Fredholm determinants related to gap probabilities, not for the matrix ones describing \emph{joint} gap probabilities for multi--time processes. It is an open problem to develop the integrable theory for the matrix Fredholm case and hence derive PDEs for multi--time joint gap probabilities (see also \cite{BertolaCafasso3}, where it has been proven that the multi--time Airy and Pearcey kernels are integrable kernels in the sense of Its--Izergin--Korepin--Slavnov \cite{IIKS}). An example of such PDEs, to be derived for the Pearcey process with inliers, can be found in \cite{ADvMVa}, Conjecture 1.5. 
	\item Since the PDEs derived ultimately come from the KP hierarchy, one might expect that they have a ``Painlev\'e property'' in a suitably general sense; for example, the solutions should have only poles along a movable divisor in the space of parameters. A first step in this study has been done in \cite{BertolaCafasso1}. Indeed, in that paper, the equations (\ref{intro10}) and (\ref{intro11}) have been re--derived (together with an additional one) as compatibility conditions of an isomonodromic system of ODEs. It is a beautiful open question to develop a theory establishing a  general connection between the equations we derived here, the theory of isomonodromic deformations and related Painlev\'e properties.
	\item The question of appropriate initial conditions for such PDEs is also an open question and might involve asymptotic solutions rather then analytic solutions of such PDEs. This question is not unrelated to the second open question.
	\item According to a well--established physics literature (see \cite{BergereEynard} and references therein) a large class of kernels (the so--called $(p,q)$--kernels) should appear considering critical phenomena in two--matrix models. These kernels should be related to Gel'fand--Dickey hierarchies supplemented with appropriate Virasoro constraints, even if no general expression is given yet (see also \cite{CIK} and references therein for the $(2,q)$--case). It would be extremely interesting to prove in full generality that the kernels (\ref{intro1}) we studied here, given an appropriate choice of the contours, correspond indeed to the $(p,1)$--kernels, as it is the case for $p=2,3$. We also mention that, using the method developed in this paper, in \cite{ACvM3} we have been able to derive PDEs for some kernels generalizing (\ref{intro1}) and including the ones studied in \cite{CIK}.

\end{enumerate}

\section{KP Theory revisited}

In this section we provide the basic tools, which will be needed, (see for example \cite{DJKM1,DJKM2,ASvM, ASvM2, solitons}) from KP theory; in particular $\tau$--function's formalism and wave functions, Grassmannian theory and additional symmetries, and also p--reduced (Gelfand--Dickey) and $k$--vector p--reduced KP hierarchies. These latter introduced by Krichever \cite{KrKP} and more specifically we will use some results due to Helminck and van de Leur \cite{HevdLcKP, vdLcKP}. We shall just give the basic facts necessary for the subsequent chapters.\\
Let $q=q(x,y,t)$ be a function of three variables; the KP equation describing shallow water waves in $\R^2$, namely
\be
	q_{xxxx}+12q_{x}^2+12qq_{xx}+3q_{yy}-4q_{xt} = 0,
\nonumber\label{KP}\ee
is actually part of a hierarchy of commuting PDE flows for $$q := \partial_x^2 \log\tau(t_1,t_2,t_3,t_4,\ldots)$$ where we should identify\footnote{actually in the sequel we prefer to keep $t_1$ and $x$ as two distinct variables; the identification is justified since $\tau$ always depends just on the sum $x+t_1$.}
$$t_1\equiv x, t_2 \equiv y, t_3 \equiv t$$ and the other ``times'' corresponds to higher order flows of the hierarchy. The whole set of equations satisfied by the hierarchy can be written, using the function $\tau({\bf t}):=\tau(t_1,t_2,t_3,t_4,\ldots)$, as the famous bilinear identity
\be
	\oint_{\infty} \tau({\bf t}-[z^{-1}])\tau({\bf t'}+[z^{-1}]){\rm e}^{\sum_i^{\infty}(t_i-t'_i)z^i} = 0.
\label{BHE}\ee
In the expression above we denoted $[z] := (z,z^2/2,z^3/3,\ldots)$ (and similarly for $z^{-1}$); the  equations of the hierarchy are obtained expanding the integrand as a formal Laurent series about $z^{-1} = 0$ and then taking the (formal) residue about $z = \infty$.   

The hierarchy can also be written as a commuting set of Lax equations for a $({x,\mathbf t})$--evolving pseudo--differential operator
\be
	L^+=L^+(x,{\mathbf t}) := D + \sum_{j=1}^{\infty} a_j(x,{\mathbf t})D^{-j},
\ee
where $D:=\partial_x$ (the reason for the notation $L^+$ instead of the classic $L$ will be clear later)
; namely
\be
\frac{\partial}{\partial t_i} L^+ = \left[\left(L^{+i}\right)_+, L^+\right], \quad i\in\Z_+ 
\label{primolax}\ee
where, for an arbitrary (formal) pseudo--differential operator $H:=\sum_{i\in\Z}h_i D^i$, we denote 
$$(H)_+:= \sum_{i\geq 0} h_iD^i,\quad\quad H_-:=H-H_+$$
(note also that from the equation (\ref{primolax}) we deduce $\dis\frac{\partial}{\partial t_1} = D$).\\

In order to derive the Lax formulation from the bilinear identity (\ref{BHE}) it is necessary to define a (suitably normalized) eigenfunction $\Psi^+$ of $L^+$, itself defined by a pseudo--differential operator $W$ introduced below. Indeed setting
\be
	\overline{\mathbf t} := {\mathbf t} + xe_1 = (x+t_1,t_2,t_3,\ldots),\quad 
	\partial_{\mathbf t} :=\left(\frac{\partial}{\partial t_1}, \frac{1}{2}\frac{\partial}{\partial t_2}, \frac{1}{3}\frac{\partial}{\partial t_3},\ldots\right)
\ee  
and denoting as before with  $p_i({\mathbf t})$ the classical Schur polynomials such that

\be\label{Schur}
	{\rm e}^{\sum_{i=1}^\infty t_iz^i}  = \sum_{i=0}^\infty z^i p_i({\mathbf t}),
\ee
we define  the wave operator $W=W(\bar{\mathbf t})$
\be
	W:=\frac{\tau(\bar{\mathbf t}-[D^{-1}])}{\tau(\bar{\mathbf t})}{\rm e}^{\sum_1^{\iy}t_iD^i}= 
	\left(\sum^{\iy}_{j=0}\frac{p_j(-\pl_{\mathbf t})\tau(\bar{\mathbf t})}{\tau(\bar{\mathbf t})}D^{- j}\right){\rm e}^{\sum_1^{\iy}t_iD^i}.
\label{waveoperator}\ee
Denoting by $H^*$ the formal adjoint of a given pseudo--differential operator $H$ (the formal adjoint acts through the formula $(a(x)D^j)^*:= (-D)^ja(x),\; j\in\Z$ and linearity), we can define the wave function $\Psi^+$ and Lax operators $L^+,M^+$ together with their adjoints (denoted with the minus sign)

\be
\begin{array}{lll}
L^+:=WDW^{-1},&M^+:=WxW^{-1},&\Psi^+: =We^{xz}\\
\\
L^-:=(W^{-1})^*(-D)W^*,&M^-:=(W^{-1})^*xW^*,&\Psi^-:=(W^{-1})^*e^{- xz}.
\end{array}
\label{7}\ee
One should think of $L^\pm$ and $M^\pm$ as the ``dressing'' of $\pm D$ and $x$ while the wave functions $\Psi^\pm$ are the ``dressing'' of ${\rm e}^{\pm xz}$; this leading to the following relations, which result from ``undressing'' the operators and functions, to wit:

\be
\begin{array}{lll}
L^\pm\Psi^\pm=z\Psi^\pm,&M^\pm\Psi^\pm =\pm\dis\frac{\pl}{\pl z}\Psi^\pm,&[L^\pm,M^\pm]=\pm1.
\end{array}
\label{8}\ee
Also the definition of the wave operator $W$ (eq. (\ref{waveoperator})) together with (\ref{7}) gives back the celebrated Sato's formula expressing the wave functions in terms of the tau function, namely
\be
	\Psi^\pm(\bar{\mathbf t};z) := {\rm e}^{\pm xz \pm \sum_{i=1}^\infty t_iz^i}\dfrac{\tau\left(\bar{\mathbf t}\mp [z^{-1}]\right)}{\tau(\bar{\mathbf t})}.
\label{SatoFormula}\ee
 It is a well known result that Sato's equation (\ref{SatoFormula}), together with the bilinear identity (\ref{BHE}), gives the following deformation equations for the wave functions \footnote{Here and below $L^{-i}$ means $(L^-)^i$ and not $L$ to the power $-i$ (on the other hand we did not introduce any operator $L$).}
\be
	\frac{\pl}{\pl t_i}\Psi^\pm =\pm(L^{\pm i})_+\Psi^\pm, \quad i\in\Z_+,\label{9'}
\ee
 and these equations, finally, give as compatibility conditions the Lax equations:
\be
 \frac{\pl}{\pl t_i}L^\pm=\left[\pm(L^{\pm i})_ +,L^\pm\right] \quad i \in \Z_+.
\label{9}\ee
As a side remark we want to point out that, till now, the pseudo--differential operators $M^\pm$ did not play any role. As a matter of fact they are not useful to define the hierarchy, but rather to define some additional symmetries which will be discussed later.\\

Sato observed that the KP flows linearize on an infinite--dimensional Grassmannian $\mathrm{Gr}$ whose elements are linear spaces $\WR$ of formal Laurent series, i.e. typically we have\footnote{As the expert readers will have noticed, we are just describing the so--called ``big cell'' of the Sato's Grassmannian, which is enough for our purposes.}
\be
	\WR = \mathrm{span}\left\{\varphi_n(z) = z^n + \sum_{-\infty<i\leq n-1} a_{ni} z^i,\;a_{ni}\in\BC\right\}_{n=0}^\infty .
\label{basis}\ee
More precisely, associating to any KP solution, through its wave functions, two subspaces $\WR^\pm$ through the formula
\be
\begin{array}{lll}
\WR^\pm&:=& {\rm span}_{i\geq  
0}   \{D^i\Psi^\pm(x,0;z)\},
\end{array}
\label{11}\ee
we get that the linear spaces $\WR^\pm$ are ${\mathbf t}$--deformed by the KP flows via
$$\WR^\pm({\mathbf t}) = {\rm e}^{\mp\sum_{i=1}^\infty t_iz^i}\WR^\pm.$$
In fact the subspaces $\WR^\pm$ defined above are an alternative way of characterizing the KP data encoded in the tau function (or also in the wave function). In particular invariance properties of the KP solutions are sometimes more easily characterized in terms of $\WR^\pm$. To this end consider the mapping from $z$--operators $A$ to $x$--operators $\mathcal P_A$ given by  

\be
\begin{array}{lll}
A^+\Psi^+&:=&\dis\sum_{-\iy <i<\iy}\sum_{j\geq 0}c_{ij}z^i\left(\dis\frac{\pl}{\pl  
z}\right)^j\Psi^+=\sum_{i,j}c_{ij}(M^{+})^j(L^{+})^i\Psi^+ =:\PR_{A^+}^+\Psi^+,\\
\\
A^-\Psi^-&:=&\dis\sum_{-\iy <i<\iy}\sum_{j\geq 0}c_{ij}\left(-\dis\frac{\pl} 
{\pl z}\right)^j z^i\Psi^- =\sum_{i,j}c_{ij}(L^{-})^i(M^{-})^j\Psi^-=:\PR_{A^-}^- 
\Psi^-,
\end{array}
\label{10}\ee
where the equalities are proved using (\ref{8}). The following identical statements follow from (\ref{10}) and relate the invariance properties of $\WR^\pm$ with regard to differential operators in $z$ to properties of the associated differential operator in $x$,
\be
A^+\WR^+\subset \WR^+\DF \PR^+_{A^+}=(\PR^+_{A^+})_+ \DF A^-\WR^-\subset \WR^-.
\label{12}\ee
In particular the points in the Grassmmannian satisfying (\ref{12}) for $A^+=z^2$ are the points associated to the KdV hierarchy, considered as a reduction of the KP one (more details will be given in the following section). It should be noted that Sato's Grassmannian, differently from the Segal--Wilson one \cite{SW}, is constructed with formal series, not analytical or asymptotics ones. Hence all the relevant computations are done with formal series, and in particular the basis in (\ref{basis}) are well defined as formal series while the tau function $\tau({\mathbf t})$ is a formal infinite series composed of monomials in the $t_i$, computable from the formal basis; in the sequel we will take full advantage of that. Nevertheless, for some particular subspaces $\WR$ and some particular values of time parameters, formal computations acquire a more concrete meaning as analytical and/or asymptotics series, (on the other hand it is worth recalling that the Segal--Wilson Grassmannian is strictly contained in the Sato one).

\subsection{p--reduced KP--hierarchies}

In this section we recall the definition (in this specific context of Sato's Grassmannian theory) of Gel'fand--Dickey hierarchies which we shall refer to as the $p$--reduced KP hierarchies. First we need the following well known lemma:
\begin{lemma}\label{equiv}
	Let $p$ be a non--negative integer. Given a point $\WR^+$ in the Sato's Grassmanian and let $L^+$ and $\tau$ be the Lax operator and the tau function of the corresponding solution of the KP hierarchy. The following conditions are equivalent and are conserved along the KP flows.
	\begin{enumerate}
		\item $z^p\WR^+\subseteq \WR^+$
		\item $L^{+p} = (L^{+p})_+$
		\item $\tau$ does not depend on $t_{np}$ for any $n \geq 1$, modulo a removable factor of the form ${\rm e}^{\sum_{i=1}^\infty c_nt_{np}}$.
	\end{enumerate}
\end{lemma}
\proof
The equivalence between 1. and 2. is given by (\ref{12}) while the one between 2. and 3. can be proven observing that, if $L^{+p} = (L^{+p})_+$, then we have
$$\frac{\partial}{\partial t_{np}} \Psi^+ = z^{np}\Psi^+\quad\forall\; n\geq 1 $$ and so 3. is equivalent to 2. using Sato's formula (\ref{SatoFormula}). The fact that the three conditions are conserved along KP flows is proven by observing that the first one is obviously conserved along the KP flows, since $\WR^+({\mathbf t}) = {\rm e}^{-\sum_{i=1}^\infty t_iz^i}\WR^+$ in Sato's Grassmannian.\qed

\begin{definition}
The p--reduced KP--hierarchy is the KP--hierarchy supplemented with one of the (equivalent) conditions 1. 2. or 3. of Lemma \ref{equiv}. 
\end{definition}
In the case of p--reduced KP hierarchies the relevant Lax operator, rather then being $L^+$, is its $p^{th}$ power $L^{+p}$, since the latter is a differential operator. Indeed, upon using the Lax formulation of the KP hierarchy and setting

$$\LR^\pm := L^{\pm p},$$
one finds equations
$$\LR^\pm \Psi^\pm = z^p \Psi^\pm,\quad \frac{\pl \Psi^{\pm}}{\pl t_i}=(\pm\LR^{\pm{i}/{p}})_+\Psi,\quad i\in\Z_+$$ 
and their compatibility conditions give the Lax formulation of the p--reduced KP hierarchy:
\be
	\frac{\pl\LR^{\pm}}{\pl  t_i}=\bigl [(\pm\LR^{\pm{i}/{p}})_+,\LR \bigr],\quad i\in\Z_+.
\label{pKP}\ee
In the sequel we are interested in some particular solutions of p--reduced KP hierarchies satisfying an additional invariance property. Let's start defining a $z$--operator 
\be
	\AR_p^\pm(z) := z\pm\dis\frac{1}{pz^p}\left(z\dis\frac{\pl}{\pl z}-\dis\frac{p-1} {2}\right);
\label{duesedici}\ee
it is easy to check that $\AR_p^\pm$ satisfies the following condition:
$$[\AR_p^\pm(z), z^p] = \pm 1.$$

\begin{theorem} \label{theo1}
%
The invariance conditions
\be\begin{aligned}
& z^p\WR^+ \subset \WR^+,\quad \AR^+_p\WR^+\subset \WR^+,\\
 \end{aligned} \label{16}\ee
determine uniquely a plane $\WR^+_p\in \mathrm{Gr}$, which moreover uniquely determines $\WR^-_p\in\mathrm{Gr}$ by the relations
 \be
z^p\WR^-\subset \WR^-,\quad\AR_p^- \WR^-\subset \WR^- . \label{17} \ee
The $\WR^\pm_p$ are linearly generated by the eigenfunctions $\vp_p^\pm$ of the operators $\AR_p^{\pm}$, namely:
\be
\WR^\pm_p={\rm span}_{i\geq 0}\{(\AR_p^\pm)^i\vp_p^\pm\}
\label{18}\ee
with
 \be (\AR_p^{\pm})^p\vp_p^\pm=z^p\vp_p^\pm,\quad\vp_p^\pm(z)=1+\sum_1^{\iy}\frac{a_i^\pm}{z^i};  \label{22}\ee
the latter uniquely determines $\varphi_p^{\pm}(z)$.\\
The corresponding p--reduced KP wave functions $\Psi^\pm_p(x,0;z)$ are then uniquely specified by
\be
\begin{array}{ll}
\Psi^\pm_p(0,0;z)=\vp_p^\pm(z)\\
\\
\AR_p^\pm(z)\Psi^\pm_p(x,0;z)=\pm\dis\frac{\pl}{\pl x}\Psi^\pm_p(x,0;z).
\end{array}
  \label{19}\ee
\end{theorem}
Parts of this theorem appeared in \cite{KacSchwarz} and \cite{ASvM}, but perhaps with sketchy proofs for the case $p > 2$. We will see in the sequel that these particular solutions of KP, determined either by $\WR^\pm_p, \varphi^\pm_p$ or $\Psi^\pm_p(x,0;z)$, can be made fairly explicit (see Theorems \ref{cor:1},\ref{3'} and Lemma \ref{lemma8}).\\
To prove the theorem above it clearly suffices to prove the following lemmas.
\begin{lemma}\label{lemma1} The condition
\be
z^p\WR^+\subset \WR^+,\quad\AR_p^+\WR^+\subset \WR^+,
  \label{20}\ee
uniquely determines $\WR^+$ together with its adjoint $\WR^-$, and these are specified by (\ref{18}), with $\vp_p^\pm(z)$ determined uniquely by the differential equations and the series (\ref{22}).
%
%
 \end{lemma}


 \begin{lemma}\label{lemma3} The wave functions $\Psi^\pm_p(x,0;z)$ going with $\WR^\pm_p$ of (\ref{20}) satisfy the equations (\ref{19}) and are uniquely determined by them. \end{lemma}

\bigbreak

 \noindent{\it Proof of Lemma \ref{lemma1}:} \,  Note that since
 $$
 \AR^+_p \WR^+\subset \WR^+,\quad \AR^+_p=z+{\bf O}\left(\frac{1}{z^p}\right),\quad \WR^+\in \mathrm{Gr},
 $$
 then for some $\vp_p^+(z)=1+\dis\sum_{i=1}^{\iy}\dis\frac{a_i^+}{z^i}$, we have that
 $$
 \WR^+={\rm span}\left\{(\AR_p^+)^i\vp_p^+=z^i\left(1+{\bf O}\left(\frac{1}{z}\right)\right)\right\};
 $$
 but since  $z^p\WR^+\subset \WR^+$, $z^p\vp_p^+(z)=z^p\left(1+{\bf O}\left(\frac{1}{z}\right)\right)$, we must have, for some constants $c_0,c_1,...,c_{p-2}$, that
 $$
 z^p\vp_p^+=\left((\AR_p^+)^p+c_{p-1}(\AR_p^+)^{p-1}+c_{p-2}(\AR_p^+)^{p-2}+...+c_0\right)\vp^+_p,
$$
and so we just must show all the $c_i=0$. Since\footnote{Here $\varepsilon_k(\cdot)$ stands for a differential operator having the property that $\varepsilon_k(z^i)=\sum_{-\infty \leq \ell \leq i+k} a_{\ell}^{(i)}z^\ell,\;-\infty < i < \infty$, the $a_{\ell}^{(i)}$ being constants.} $\AR_p^+ =z+\vr_{-p}$, by induction $(\AR_p^+)^{p-j}=z^{p-j}+\vr_{-j-1}$, and so

\bean
0&\equiv&\left((\AR_p^+)^p+c_{p-1}(\AR_p^+)^{p-1}+c_{p-2}(\AR_p^+)^{p-2}+...+c_0-z^p\right)\vp^+_p(z)\\
\\
&=&(\vr_{-1}+c_{p-1}(z^{p-1}+\vr_{-2})+c_{p-2}(z^{p-2}+\vr_{-3})+...+c_1(z+\vr_{-p})+c_0)\\
& &\hspace*{8cm}\left(1+\sum_1^{\iy}\frac{a_i^+}{z}\right).
\eean
So we first conclude $c_{p-1}=0$, since the left hand side has no matching power of $z^{p-1}$, then $c_{p-2}=0$, and so inductively, $c_0=0$.
Finally (\ref{12}), (\ref{20}) implies $z^p\WR^-\subset \WR^-$,  $\AR_p^-\WR^-\subset \WR^-$ and we can repeat the above arguments for $\WR^-$, concluding the proof of Lemma \ref{lemma1}, except that it remains to be shown that (\ref{22}) determines $\varphi_p^\pm(z)$ uniquely. 
To do so, from the above $(\AR^+_p)^p=z^p+\vr_{-1}$, and more precisely
 $$
 ((\AR_p^+)^p-z^p)=\frac{\pl}{\pl z}+\dt_{-2}+\dt_{-3}+...+\dt_{-p^2},
 $$
 where $\dt_k$ is an operator such that deg$(\dt_k(z^i))=i+k$, for $-\iy <i<\iy$. We find from (\ref{22}) 
 \bean
 0&\equiv& (\left(\AR_p^+\right)^p-z^p)\vp^+_p(z)=\left(\frac{\pl}{\pl z}+\dt_{-2}+\dt_{-3}+\ldots\right)\left(1+\frac{a_1^+}{z}+\frac{a_2^+}{z^2}+\ldots\right)\\
 \\
 &=&\left(-\frac{a_1^+}{z^2}+\dt_{-2}(1)\right)+\left(-\frac{2a_2^+}{z^3}+\dt_{-2}\left(\frac{a_1^+}{z}\right)+\dt_{-3}(1)\right)\\
 \\
 & &+\left(-\frac{3a_3^+}{z^4}+\dt_{-2}\left(\frac{a_2^+}{z^2}\right)+\dt_{-3}\left(\frac{a_1^+}{z}\right)+\dt_{-4}(1)\right)+\ldots =:\Gamma_{-2}+\Gamma_{-3}+\Gamma_{-4}+\ldots,
 \eean
 with degree $\Gamma_i =i$. Since all the terms $\Gamma_i$ individually must be 0, we recursively determine $a_j^+$ from $\Gamma_{j-3}=0$, $j\geq 1$, and similarly for $\varphi_p^-(z)$.\qed

 \bigbreak

 \noindent{\it Proof of Lemma \ref{lemma3}:} \,  From $\WR^+_p\in \mathrm{Gr}$, and comparing (\ref{18}) and (\ref{20}), namely $ {\rm span}_{i\geq  
0}   \{D^i\Psi^+(x,0;z)\}={\rm span}_{i\geq 0}\{(\AR_p^+)^i\vp_p^+\}$, and observing both $\vp_p^+(z)$ and $\Psi^+(0,0;z)$ are $\left(1+{\bf O}\left(\frac{1}{z}\right)\right)$ (from (\ref{waveoperator}),(\ref{7}) and $\tau(x,0,0,\ldots)\neq 0$) conclude $\Psi^+(0,0;z)_p=\vp_p^+(z)$. Moreover, from (\ref{10}), (\ref{20}) and (\ref{12}) conclude, in particular, that
 $$
 \AR_p^+(z)\Psi^+_p(x,0;z)=(\PR^+_{\AR^+_p}\Psi^+_p(x,{\mathbf t};z))\Big\vert_{{\mathbf t}=0}=(\PR^+_{\AR^+_p}(x,0;z))_+\Psi^+_p(x,0;z).
 $$
 Now denote with $L_p^+,M_p^+$ the two Lax operators corresponding to this particular solution of the p--reduced KP--hierarchy; from (\ref{10}) and (\ref{duesedici}) we find
 \be
 (\PR^+_{\AR^+_p})_+\big\vert_{{\mathbf t}=0}=\left(L^+_p +\frac{1}{p}(M^+_p L^+_p -\frac{p-1}{2})(L^+_p)^{-p}\right)_+\big\vert_{{\mathbf t}=0}=(L^+_p)_+\big\vert_{{\mathbf t}=0}=\frac{\pl}{\pl x}
\label{25} \ee
 and thus
 $$
 \AR_p^+(z)\Psi_p^+(x,0;z)=\frac{\pl}{\pl x}\Psi_p^+(x,0;z),\quad\Psi^+_p(0,0;z)=\vp_p^+(z),
 $$
 the latter being a nonsingular first order PDE with given initial condition, which completely determines $\Psi_p^+(x,0;z)$, as claimed. Meanwhile from (\ref{10}), (\ref{12}), (\ref{20}) and (\ref{duesedici}), conclude that
 $$
 \AR^-_p(z)\Psi^-_p(x,0;z)=(\PR^-_{\AR_p^-}\Psi^-_p(x,{\mathbf t};z))\Big\vert_{{\mathbf t}=0}=-\frac{\pl}{\pl x}\Psi^-_p(x,0;z),
 $$
and then as before $\Psi^-_p(0,0;z)=\vp^-_p(z)$, with $\Psi^-_p(x,0;z)$ uniquely determined as before, concluding the proof of Lemma \ref{lemma3}. \qed

 \subsection{Airy-like integrals and asymptotics}

In this section we show how to explicitly construct (in Theorem \ref{cor:1}) the wave functions $\Psi^\pm_p(x,0;z)$ described in Theorem \ref{theo1}. Their asymptotics is given in Theorem \ref{3'} together with some examples in Corollary \ref{contornigiusti}.\\
 As an application of Theorem \ref{theo1}, consider the functions
 \be
\begin{array}{lll}
&\Phi_p^\pm(u):=\dis\sqrt{\dis\frac{\pm p}{2\pi}}\dis\int_{\Gamma_p^\pm}{\rm e}^{\mp\frac{y^{p+1}}{p+1}\pm uy}dy,~~
\end{array}
\label{27}\ee
with $\Gamma_p^\pm$ being a union of lines in $\BC$ through the origin, picked so that (\ref{27}) makes sense, and such that there exists regions $\DR_p^\pm$ which are unions of sectors about the origin in which $\Phi_p^\pm$ have the following asymptotic behavior (see Theorem \ref{3'}) in $\DR_p^\pm$:
\be
\begin{aligned}
\Phi_p^\pm(z^p)&= z^{- \frac{p-1}{2} }{\rm e}^{\frac{\pm p}{p+1}z^{p+1}}\Bigl(1+\sum_1^{\iy} {a_i^\pm}{z^{-i}}\Bigr),~~~\mbox{for}~
  z\in\DR_p^\pm.
\end{aligned}
\label{28}\ee

\begin{theorem}\label{cor:1}
The wave functions $\Psi_p^\pm(x,0;z)$ of Theorem \ref{theo1} are given by the following formula:
\be  
\Psi_p^\pm(x,0;z) = z^{\frac{p-1}2}{\rm e}^{\mp \frac{p}{p+1}z^{p+1}}\Phi^\pm(x+z^p).
\label{40''bis}\ee
Moreover they satisfy the spectral equation
\be
 	\LR_p^\pm(x,0)\Psi_{p}^{\pm}(x,0;z) := ((\pm D)^p-x)\Psi_{p}^{\pm}(x,0;z)=z^p\Psi_{p}^{\pm}(x,0;z).
  \label{26}\ee
\end{theorem}

\proof Using Theorem \ref{theo1} we simply have to prove equations (\ref{22}) and (\ref{19}). Note that by our choice of $\Gamma_p^\pm$, $c_p:=\sqrt{\pm 2\pi/p}$, the functions $\Phi_p^\pm$ are solutions of the spectral problem:
\be\begin{aligned}
\left(\mp\left(\pm\frac{d}{du}\right)^p\pm u\right)\Phi_p^\pm(u)&=\frac{1}{c_p}\int_{\Gamma_p^\pm} (\mp y^p\pm u){\rm e}^{\mp\frac{y^{p+1}}{p+1} \pm uy} dy\\
&=\frac{1}{c_p}\int_{\Gamma_p^\pm}\frac{d}{dy}\left({\rm e}^{\mp \frac{y^{p+1}}{p+1}\pm uy}\right)dy=0,
\end{aligned}\label{34}\ee
which implies (\ref{26}). Note that (\ref{28}) shows that $\Psi_p^\pm(0,0;z) = 1 + \sum_{i = 1}^\infty a_i^\pm z^{-i}$ and so to prove the first relation in (\ref{19}) we need to prove the following equation:

\be
(\AR_p^\pm)^p(z)\Psi_p^\pm(0,0;z) = z^p\Psi_p^\pm(0,0;z).
\label{aez}\ee

 To that end, observe, using (\ref{duesedici}) and (\ref{28}) that
\bean
\frac{d}{du}\Phi_p^\pm(u)\Big\vert_{u=z^p}=\frac{1}{pz^{p-1}}\frac{d}{dz}\Phi_p^\pm(z^p)
&=&\frac{1}{pz^{p-1}}\frac{d}{dz}\left(z^{-\frac{p-1}{2}}{\rm e}^{\pm \frac{p}{p+1}z^{p+1}}\Psi_p^\pm(0,0;z)\right)\\
\\
&=&z^{-\frac{p-1}{2}}e^{\pm \frac{p}{p+1}z^{p+1}}\AR_p^\pm(z)\Psi_p^\pm(0,0;z),
\eean
and so conclude that
\be
\pm\frac{d}{du}\Phi_p^\pm(u)\Big\vert_{u=z^p}=z^{-\frac{p-1}{2}}{\rm e}^{\pm \frac{p}{p+1}z^{p+1}}(\AR_p^\pm(z)\Psi_p^\pm(0,0;z)).
\label{2.29'}\ee
Repeating the argument $p$--times and using (\ref{28}) and (\ref{34}) conclude that
$$
\begin{aligned}
z^p\left(z^{-(\frac{p-1}{2})}{\rm e}^{\pm \frac{p}{p+1}z^{p+1}}\Psi_p^\pm(0,0;z)\right)&=(u\Phi_p^\pm(u))\big\vert_{u=z^p}= \left(\pm\frac{d}{du}\right)^p\Phi_p(u)\big\vert_{u=z^p}\\
&= z^{-(\frac{p-1}{2})}{\rm e}^{\pm \frac{p}{p+1}z^{p+1}}\left((\AR_p^\pm)^p(z)\Psi_p^\pm(0,0;z)\right);
\end{aligned}
$$
and this yields (\ref{aez}), which yields (\ref{22}) and the first equation in (\ref{19}), while the second equation is a consequence of the calculation (\ref{2.29'}) with $u \mapsto x + z^p$ instead of $z^p$. Theorem \ref{3'} yields (\ref{28}), which was used to derive the series expansion for $\Psi_p^\pm(0,0,z) = \varphi_p^\pm(z)$. \qed 
Prior to proving that the formal series in (\ref{22}) are actually asymptotic expansions, we remind the reader of a well known theorem concerning asymptotic expansion of solutions of holomorphic differential equations (see Theorem 19.1 of \cite{Wasow}), here slightly rephrased for our applications. For $C$ a constant matrix (not necessarily diagonalizable) written in Jordan form as $C=\mathcal U D \mathcal U^{-1}$, we denote as usual
$z^C:={\rm e}^{C\log z} = \mathcal U z^D \mathcal U^{-1}$ and thus $z^C$ can be multivalued, depending on $D$. We now have

\begin{theorem} (Wasow)  Let $A(z)$ be an $n\times n$ matrix function, holomorphic for $|z|\geq |z_0|$, $z\in S$, where $S$ is an open sector with vertex at the origin. Assume that $A(z)$ possesses an asymptotic series ``in powers of $z^{-1}$", i.e. $A(z)=\displaystyle{\sum_{i\leq\ell}}A_iz^i$. Then corresponding to every sufficiently narrow open subsector $S$, the ODE: \newline $Y'=A(z)Y$ possesses a fundamental matrix solution in $S$ of the form

$$
Y(z)=\left(\sum_{i\geq 0}\frac{B_i}{(z^{1/p})^i}\right)z^C \diag(e^{P_1(z^{1/p})},...,e^{P_n(z^{1/p})}),
$$
with $p$ a positive integer, the $P_i$ polynomials in their arguments, $C$ a constant matrix, and the sum is an asymptotic series. The solution depends on the arbitrary choice of the branch of $z^{1/p}$.
\end{theorem}

The following theorem proves that the formal expansions of (\ref{22}) are actually asymptotic expansions. It is worth noting that the expansion are independent of the sector they occur in.  

\begin{figure}
\resizebox{1\textwidth}{!}{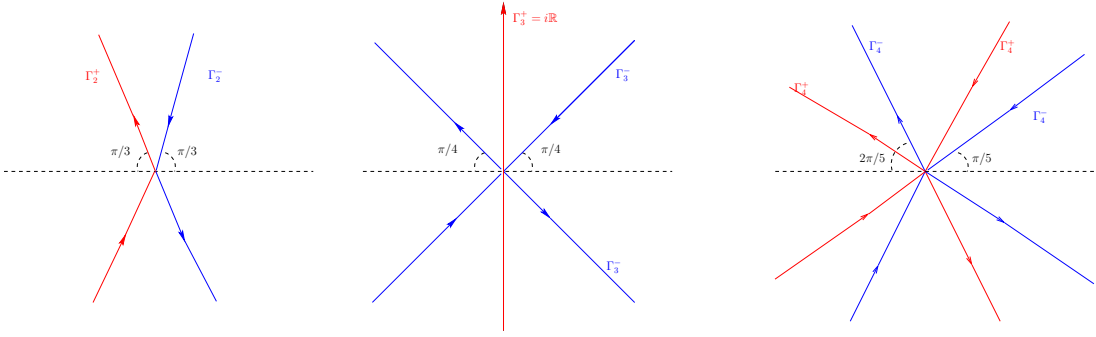}
\caption{A possible choice of the integration contours for $p=2,3,4$.}
\label{Contours}
\end{figure}

\begin{theorem} \label{3'}(Asymptotics for general $p$)  Consider the integrals
\be
\Phi_p^\pm(u)=\sqrt{\pm\frac{p}{2\pi}}\int_{\Gamma_p^\pm}{\rm e}^{\mp \frac{y^{p+1}}{p+1} \pm uy}dy\label{40}\ee

over contours (see Figure \ref{Contours}) $\Gamma_p^\pm$, with $\om =e^{\frac{\pi i}{p+1}}$,
\be
\Gamma_p^+ \subset \left\{{\cal C}( \om^{2j}),~~~~j\leq \Big[\frac{p+1}{2}\Big]\right\},~~~\Gamma_p^- \subset \left\{{\cal C}( \om^{2j+1}),~~j\leq \Big[\frac{p}{2}\Big]\right\}.
\label{41rays}\ee
Then $\Phi_p^\pm$ have expansions (independent from the sector) of the form
\be
\Phi_p^\pm(z^p)=z^{-(\frac{p-1}{2})}{\rm e}^{\pm\frac{p}{p+1}z^{p+1}}\left(1+\sum_{i\geq [\frac{p+2}{2}]}\frac{a_i^\pm}{z^i}\right)
\label{42}\ee
for sectors of size $\frac{2\pi}{p(p+1)}$ centered about the rays  
\be
{\cal C}(\om^{2j})\subset\Gamma_p^+,\quad {\cal C}(\om^{(2j+1)})\subset\Gamma_p^-.
\label{43}\ee
The contours are all oriented counter-clockwise with the contour\\ ${\cal C}(\omega^\ell)=\BR^+\cdot\omega^\ell+\BR^+\cdot\overline{\omega}^\ell$ as a point-set.
\end{theorem}

\proof Considering a specific contour ${\cal C}(\omega^{2j})\subseteq\Gamma_p^+$ and the sector of angle size $\frac{2\pi}{p(p+1)}$ centered about this ray; the points along a ray within  this sector can be parametrized by $\om^{2j}{\rm e}^{i\theta}z$, with $z\in\BR^+$, and $|\theta |<\frac{\pi}{p(p+1)}$.
Then, setting $y\mapsto yz$, $z\in\BR^+$ in the integrand in (\ref{40}), one obtains
\be
\Phi_p^+((\om^{2j}{\rm e}^{i\theta}z)^p)=\sqrt{\frac{p}{2\pi}}z\int_{\Gamma_p}{\rm e}^{V(y)z^{p+1}}dy,
 ~~\mbox{with}~~V(y)=-\frac{y^{p+1}}{p+1}+{\rm e}^{ip\theta}\om^{2jp}y.
\label{43a}\ee
Employing the saddle point method, first note
$$
V'(y_0)=0\Longrightarrow y_0={\rm e}^{i\theta}\om^{2j}\cdot \{1,\gamma,...,\gamma^{p-1}\},\quad\gamma ={\rm e}^{\frac{2\pi i}{p}}.
$$
Then, using $\om ={\rm e}^{\frac{\pi i}{p+1}}$,
\bean
V(y_0)z^{p+1}=
 \frac{p}{p+1}(z~{\rm e}^{i\theta})^{p+1} \{1,\gamma,...,\gamma^{\ell},...,\gamma^{p-1}\},
\eean
and since $z\in\BR^+$,

$$
\Re (V(y_0)z^{p+1})=\frac{p}{p+1}z^{p+1}\cos\left((p+1)\theta +\frac{2\pi\ell}{p}\right),\quad 0\leq\ell\leq p-1.
$$
%
Except for the prefactor $\frac{p}{p+1}z^{p+1}$, the values of $\Re (V(y_0)z^{p+1})$ are given by 
\be
\cos\left((p+1)\theta  \right),\cos\left((p+1)\theta +\frac{2\pi }{p}\right),
\ldots, \cos\left((p+1)\theta +\frac{2\pi (p-1)}{p}\right), \label{seq}
\ee
with the maximum given by the first number, namely $\cos\left((p+1)\theta  \right)$; that is for $\ell=0$. Indeed, when $\theta\geq 0$, then $\cos ((p+1)\theta)>0$, since $0\leq (p+1)\theta<\pi/2$. From there on, the sequence (\ref{seq}) goes down and then goes up again, finally up to 
$$
\cos\left((p+1)\theta +\frac{2\pi (p-1)}{p}\right)=
 \cos\left(\frac{2\pi  }{p}-(p+1)\theta \right)<\cos\left((p+1)\theta  \right),
$$
since $\pi> {2\pi  }/{p}-(p+1)\theta >(p+1)\theta $, upon using $|\theta |<\frac{\pi}{p(p+1)}$. A similar argument holds when $\theta<0$. 
%
 Thus only $y_0={\rm e}^{i\theta}\om^{2j}$, among the $p$ roots of $V'(y_0)=0$, will count in the saddle point analysis, since this point belongs to the sector centered about $\om^{2j} $. 
 So the contour ${\cal C}(\om^{2j})$ can be deformed to pick up the saddle point $y_0={\rm e}^{i\theta}\om^{2j}$. The function $V(y)$ has the following form, near $y_0$:
 $$
 V(y)=V(y_0)-\frac12 \left( (y-y_0)\sqrt{-V''(y_0)}\right)^2+O(y-y_0)^3,~~\mbox{with}~
 -V''(y_0)=py_0^{p-1},
 $$
 which leads to the new integration variable $u\in \BR$:
 $$
  y=y_0+\frac{( -V''(y_0))^{-1/2}}{z^{\frac{p+1}2}} u=y_0+\frac{p^{-1/2}y_0^{(1-p)/2}}{z^{\frac{p+1}2}} u.
 $$
  So localizing about $y_0$, compute, setting $\hat z:={\rm e}^{i\theta}\om^{2j}z $,
\bea
\lefteqn{\Phi^+_p(\hat z^p)=\Phi^+_p((\om^{2j}{\rm e}^{i\theta}z)^p)}\no\\
&=&\sqrt{\frac{p}{2\pi}}{\rm e}^{\frac{p}{p+1}(z~{\rm e}^{i\theta}\om^{2j})^{p+1}}z\int_{
\Gamma_p}{\rm e}^{-\frac{1}{2}(z^{\frac{p+1}{2}}\sqrt{p}~y_0^{\frac{(p-1)}{2}}(y-y_0))^2}\left(1
+{\bf O}\left(\frac{1}{z^{\frac{p+1}{2}}}\right)\right)dy\no\\
\no\\
&=&\sqrt{\frac{p}{2\pi}}{\rm e}^{\frac{p}{p+1}(z~{\rm e}^{i\theta}\om^{2j})^{p+1}}\frac{z\sqrt{2\pi}}{z^{\frac{p+1}{2}}\sqrt{p}~y_0^{\frac{p-1}{2}}}\left(1
+{\bf O}\left(\frac{1}{z^{\frac{p+1}{2}}}\right)\right)\no\\
\no\\
&=&(z~ {\rm e}^{i\theta}\om^{2j})^{-(\frac{p-1}{2})}e^{\frac{p}{p+1}(z~{\rm e}^{i\theta}\om^{2j})^{p+1}}\left(1
+{\bf O}\left(\frac{1}{z^{\frac{p+1}{2}}}\right)\right)\no\\
\label{2.36}\\
&=&\hat z^{-(\frac{p-1}{2})}{\rm e}^{\frac{p}{p+1}\hat z^{p+1}}  \left(1
+{\bf O}\left(\frac{1}{\hat z^{\frac{p+1}{2}}}\right)\right). 
 \no
\eea
Finally to prove (\ref{42}), we must appeal to Wasow's theorem , with our $p$ being identified with the $p$ of the theorem. Indeed by (\ref{34}) we find

$$
Y(z)=(\Phi^+_p(z),(\Phi^{+}_p)'(z),...,(\Phi_p^{+})^{(p-1)}(z))^{\top},\quad Y'=A(z)Y
$$

$$
A(z)=\left[ \begin{array}{cccccc}
0& 1 & & &{\bf O}\\
 & & & &\\
 \\
  &{\bf O}& & &1\\
  z& & & &0
\end{array} \right],
$$
yielding (\ref{42}) by Wasow's Theorem and (\ref{2.36}). 

To do the case of $\Phi^-_p$, assume ${\cal C}(\omega^{2j+1})\subseteq\Gamma_p^-$ and in (\ref{40}) compute instead $\Phi_p^-((\om^{2j+1}{\rm e}^{i\theta}z)^p)$, replacing (\ref{43a}). The function $V(y)$ now reads
$$
V(y)=\frac{y^{p+1}}{p+1}-{\rm e}^{ip\theta}(\om^{2j+1})^py,~\mbox{with}~V'(y_0)=0~\mbox{for}~y_0={\rm e}^{i\theta}\om^{2j+1}\{1,\gamma,...,\gamma^{p-1}\}
$$
Then one checks:
$$
V(y_0)z^{p+1}=\frac{p}{p+1}(z~{\rm e}^{i\theta})^{p+1}\{1,\gamma,...,\gamma^{p-1}\}
$$
and analogously the root $y_0=e^{i\theta}\om^{2j+1}$ will now dominate in the saddle point  analysis. Since $\Gamma^-_p\supset {\cal C}(\om^{2j+1})$, we can deform $\Gamma^-_p$ to pickup the saddle point $y_0=e^{i\theta}\om^{2j+1}$. The only difference now with the previous case is that, when we localize, the Gaussian has a positive sign, hence $\sqrt{\frac{-p}{2\pi}}$ appears in (\ref{40}) and $iy_0^{\frac{(1-p)}{2}}$ (instead of $y_0^{\frac{1-p}2}$) is the direction of steepest descent, concluding the proof.\qed

\begin{corollary}[Examples of Theorem \ref{3'}]\label{contornigiusti}
We can pick $\Phi_p^\pm$ for\\ $p=2,3,4$ as follows (see Figure \ref{Contours}):
\bea
\Phi_2^+(u)=\frac{1}{\sqrt{\pi}}\int_{\Gamma_2^+}{\rm e}^{-y^3/3+uy}dy, & &\Phi^-_2(v)=\frac{1}{i\sqrt{\pi}}\int_{\Gamma_2^-}{\rm e}^{y^3/3-vy}dy\\
\no\\
\Phi_3^+(u)=\sqrt{\frac{3}{2\pi}}\int^{i\iy}_{-i\iy}{\rm e}^{-y^4/4+uy}dy, & & \Phi^-_3(v)=\sqrt{\frac{3}{ 2\pi}}i\int_{\Gamma_3^-}{\rm e}^{y^4/4-vy}dy\\
\no\\
\Phi_4^+(u)=\sqrt{\frac{2}{\pi}}\int_{\Gamma_4^+} {\rm e}^{-y^5/5+uy}dy,& & \Phi^-_4(v)=\sqrt{\frac{2}{\pi}}i\int_{\Gamma_4^-} {\rm e}^{y^5/5-vy}dy
\eea
which leads to the following asymptotic behavior of (\ref{28})
$$
\Phi_p^\pm(z^p)=z^{-\frac{p-1}{2}}{\rm e}^{\pm\frac{p}{p+1}z^p}\bigl(1+\sum_1^{\iy}{a_i^\pm}{z^{-i}}\bigr)$$
for $p=2,3,4$ in regions $\DR_p^\pm$.

\end{corollary}

\subsection{$k$-Vector p-reduced KP-hierarchy}

We shall need the following theory of Helminck and van de Leur \cite{HevdLcKP,vdLcKP} generalizing (in the setting of Grassmannians) some known results for p-reduced KP to the case of the rational reductions of KP introduced by Krichever \cite{KrKP}. It is clear that the conditions below, in the Definition \ref{def}, are conserved along the KP flows,  thus leading to a reduction of the KP hierarchy.
 
\begin{definition}\label{def}
	The $k$--vector $p$--reduced KP--hierarchy is the KP--hierarchy corresponding to points $\WR$ in the Grassmannian such that 
	\be
	\mbox{\em a subspace $\WR'\subseteq \WR$ of codimension $k$ exists with $z^p\WR'\subseteq \WR$}.
	\label{k-red}\ee 
\end{definition}
It turns out that the solutions of the $k$--vector p--reduced KP-hierarchy correspond to a Lax operator of a particular form.

\begin{proposition}
	Given a solution of the $k$--vector $p$--reduced KP--hierarchy let $L$ be its Lax operator. Then the pseudo-differential operators $\LR^\pm:=(L^\pm)^p$, satisfying the Lax equations (\ref{pKP}), also satisfies
\be
	\LR^+_-=\sum_{j=1}^kq_j^+D^{-1}q_j^-
\label{90}\ee
with the functions $q_j,r_j$ flowing according to:
\be
	\frac{\pl q_j^\pm}{\pl t_n}=\pm\left(L^{\pm n}\right)_+q_j^\pm,\quad 1\leq j\leq k,\quad n\geq 1.
\label{91}\ee
\end{proposition}
The case $k=0$, namely $\WR=\WR'$, reduces to the standard p-reduced KP hierarchy; in the next section we will just use the case $k=1$, while in this section all the results are stated for the general case.
As an application of this theory, consider, for given $w_i\in \BR$,
\bea
	\Phi^\pm_{p,n}(u) &=& \dis\sqrt{\dis\frac{\pm p}{2\pi}}\dis\int_{\Gamma_p^\pm}e^{\mp \frac{y^{p+1}}{p+1} + uy}\prod^k_1(y-w_i)^{\pm n_i}dy,\label{94a}
\eea
with $\Gamma_p^\pm$ picked as in (\ref{27}) such that  there exist regions $\DR_p^\pm$, 
 unions of sectors, in which $\Phi_p^\pm$ have the following asymptotic behavior in $\DR_p^\pm$, with $n=\sum_1^k n_i$,
\be \begin{aligned}
\Phi_{p,n}^\pm(z^p)&= z^{\pm n-(\frac{p-1}{2})}{\rm e}^{\pm \frac{p}{p+1}z^{p+1}}\Bigl(1+\sum_1^{\iy}\frac{a_i^\pm}{z^i}\Bigr),~~\mbox{for}~
 z\in\DR_p^\pm.
\end{aligned}
\label{96}\ee
Therefore the functions (generalizing (\ref{40''bis}))
\be  
	\Psi_{p,n}^\pm(x,0;z) := z^{\mp n + \frac{p-1}{2}}{\rm e}^{\mp \frac{p}{p+1}z^{p+1}}\Phi^\pm_{p,n}(x+z^p) 
\label{kvpwavefunctions}\ee
behave, for $x=0$, as asymptotic series in $z^{-1}$, as
\be\label{97a}
\Psi_{p,n}^\pm(0,0;z)=1+\sum_1^{\iy}{a_i^\pm}{z^{-i}}.
\ee
Now define the operators $\AR_{p,n}^\pm$ (generalizing (\ref{duesedici}))
\be
\AR_{p,n}^\pm(z)=z\pm\frac{1}{pz^p}\left(z\frac{\pl}{\pl z}\pm n-\frac{p\!-\!1}{2}\right),~~\mbox{with}~
 [\AR_{p,n}^\pm,z^p]=\pm 1\label{97},\ee
and the polynomial 
\be
	P_k(x):=\prod_{i=1}^k(x-w_i),\quad P_0(x):=1.
\ee
We now have the analogue of Theorems \ref{theo1} and \ref{cor:1}, but for the $k$-vector p-reduced KP-hierarchy:
\begin{theorem}\label{theorem5.2} The functions $\Psi^\pm_{p,n}(x,0;z)$ in (\ref{kvpwavefunctions}) are, for $x=0$, the unique solutions of the following differential equations
\be
\begin{array}{lll}
 \quad \left[\left((\AR^\pm_{p,n})^p-z^p\right)P_k(\AR^\pm_{p,n})-P'_k(\AR^\pm_{p,n})-\dis\sum^k_{i=1}n_i\dis\prod_{j\neq i}(\AR_{p,n}^\pm-w_j)\right]\Psi^\pm_{p,n}(0,0;z)=0
 \end{array}
\label{104}\ee
supplemented with the asymptotic conditions (\ref{97a}).
They are the KP wave and adjoint wave functions at ${\mathbf t} = 0$ which correspond, in the Grassmannian settings, to the subspaces $\WR^\pm_{p,n}$ defined by
\be
\WR^\pm_{p,n}:={\rm span}\{(\AR^\pm_{p,n})^i\Psi_{p,n}^\pm(0,0;z)\}_{i\geq 0}.
\label{105}\ee
The $\WR^+_{p,n}$ satisfies the $k$--vector p--reduced condition (\ref{k-red}) in definition \ref{def}, for
\be \begin{aligned}
\WR'^\pm_{p,n}:=P_k(\AR^\pm_{p,n})\WR^\pm_{p,n}\subset \WR^\pm_{p,n}.
\end{aligned}\label{107}\ee
We also have
\be 
	\AR_{p,n}^\pm\WR^\pm_{p,n}\subset \WR^\pm_{p,n}.
\label{107bis}\ee
The Lax operators $\LR_{p,n}^\pm(x,0)$, at ${\mathbf t = 0}$, are given by the formula:
\be
	\left[(\pm D)^p -x -\sum_{i=1}^k n_i {\rm e}^{\pm x w_i}(\pm D)^{-1}{\rm e}^{\mp xw_i}\right] \Psi_{p,n}^\pm(x,0;z) = z^p \Psi_{p,n}^\pm(x,0;z),
\label{107tris}\ee
and so they satisfy (\ref{90}) and (\ref{91}) with 
\be
	q_i^\pm(x,0) = \mp\sqrt{n_i}{\rm e}^{\pm xw_i}.
\label{qandr}\ee
\end{theorem}




 \noindent{\it Proof 
  :} \, At first, check that
 \bean
 0&=&\sqrt{\frac{p}{2\pi}}\int_{\Gamma_p^+}\frac{\pl}{\pl y}\left(e^{-\frac{y^{p+1}}{p+1}+uy}\prod^k_{i=1}(y-w_i)^{n_i}\right)dy\\
 \\
 &=&\sqrt{\frac{p}{2\pi}}\int_{\Gamma_p^+}\left( -y^p+u+\sum^k_1\frac{n_i}{y-w_i} \right)   \left(e^{-\frac{y^{p+1}}{p+1}+uy}\prod^k_{i=1}(y-w_i)\right)\\
 &=&\sqrt{\frac{p}{2\pi}}\int_{\Gamma_p^+}\left[-\left(\frac{\pl}{\pl u}\right)^p +u+ \sum^k_{i=1}  n_i\left(\frac{\pl}{\pl u}-w_i\right)^{-1}\right]
  \left(e^{-\frac{y^{p+1}}{p+1}+uy}\prod^k_{i=1}(y-w_i)\right)\\
  \\
&=&  \left(-\left(\frac{\pl}{\pl u}\right)^p +u+ \sum^k_{i=1}  n_i\left(\frac{\pl}{\pl u}-w_i\right)^{-1}\right)
  \Phi_{p,n}^+(u),
  \eean
  and similarly
  \be
  0=\left(\left(-\frac{\pl}{\pl v}\right)^p-v-\sum^k_{i=1}n_i\left(-\frac{\pl}{\pl v}-w_i\right)^{-1}\right)\Phi^-_{p,n}(v) .
  \label{101}\ee
This yields 
\be
\begin{array}{lll}
	\left[P_k\left(\pm \dis\frac{\pl}{\pl u}\right)\left(\left(\pm\dis\frac{\pl}{\pl u}\right)^p-u\right)-\dis\sum^k_{i=1}n_i\dis\prod_{j\neq i}
	\left(\pm\frac{\pl}{\pl u}-w_j\right)\right]\Phi_{p,n}^\pm(u)=0,
\end{array}
\label{100}\ee
 upon acting on the above two identities with $P_k(\pm\frac{\pl}{\pl u})$ or alternately one may directly check  (\ref{100}) in the course of the above argument and reinterpret it as  (\ref{101}).\\
Equations  (we recall that $D=\frac{\pl}{\pl x}$)
 \be
\begin{array}{lll}
\left((\pm D^p)-x-\dis\sum^k_{i=1}n_i{\rm e}^{ \pm xw_i} (\pm D)^{-1} {\rm e}^{\mp xw_i}\right)\Phi_{p,n}^\pm(x+z^p)=z^p\Phi_{p,n}^\pm(x+z^p)
\end{array}
\label{102}\ee
 follow immediately, upon noting that
\be
(\pm D-w)^{-1}=e^{\pm xw}(\pm D^{-1})e^{\mp xw},
\label{103}\ee
yielding (\ref{107tris}) and (\ref{qandr}).\\
Also observe that (\ref{102}) is reminiscent of (\ref{26}) and is an example of (\ref{90}), with $(L^{+p})_+=D^p-x$.
In the same way as in Theorem \ref{cor:1}, deduce
 \be
 \begin{array}{lll}
 	\left(\pm \dis\frac{\pl}{\pl u}\right)^{\ell}\Phi_{p,n}^\pm(u)\Big\vert_{u=z^p}&=&z^{\pm n-(\frac{p-1}{2})}{\rm e}^{\pm \frac{p}{p+1}z^p}\left((\AR^\pm_{p,n})^{\ell}\Psi_{p,n}^\pm(0,0;z)\right),
 \end{array}
 \label{111}\ee
 and rewrite (\ref{100}) as
\be
\begin{array}{lll}
   \left[\!\left(\!\left( \pm\dis\frac{\pl}{\pl u}\right)^p-u\right)\!P_k \left( \pm\dis\frac{\pl }{\pl u}\right)-P'_k \left( \pm\dis\frac{\pl }{\pl u}\right)- \dis\sum^k_{i=1} n_i \dis\prod_{j\neq i}  \left( \pm\frac{\pl }{\pl u}-w_j\right)\!\right]\! \Phi_{p,n}^\pm(u)=0&,
\end{array}
\label{112}\ee
and thus (\ref{111}) and (\ref{112}) immediately yield (\ref{104}). Note that (\ref{104}) uniquely defines\footnote{Using $\delta_{k-1}(1/z^\ell)\neq z^k\frac{\pl}{\pl z}(1/z^\ell),\ell\geq 1$ and $\delta_{k-1}(1)=0$.
Note in Lemma \ref{lemma1} $k=0,\delta_{-1}=0$.} $\Psi_{p,n}^\pm(0,0;z)=1+\displaystyle{\sum_1^{\iy} {a_i^\pm}{z^{-i}}}$, since (\ref{104}) is of the form (see proof of Lemma \ref{lemma1} for notation)
$$
\begin{aligned}
\left(z^k\frac{\pl}{\pl z}+\dt_{k-1}+\ldots\right)\Psi_{p,n}^\pm(0,0;z)=0,
\end{aligned}
$$
and so the corresponding points in the Grassmannian $\WR^\pm_{p,n}$ are uniquely defined by (\ref{104}) and (\ref{105}), by almost the same argument as in Lemma \ref{lemma1}.

We shall now check (\ref{107}); equation (\ref{104}) implies
\be
z^pP_k(\AR_{p,n}^\pm)\Psi_{p,n}^\pm(0,0;z)\in \WR^\pm_{p,n} ,
\label{113}\ee
while for $\ell\geq 0$,
$$ \begin{aligned}
 z^pP_k(\AR_{p,n}^\pm)(\AR^\pm_{p,n})^{\ell}\Psi_{p,n}^\pm(0,0;z)&=z^p(\AR^\pm_{p,n})^{\ell}P_k(\AR_{p,n}^\pm)\Psi_{p,n}^\pm(0,0;z)
\\  
&= \bigl((\AR^\pm_{p,n})^{\ell}z^p+[z^p,(\AR^\pm_{p,n})^{\ell}]\bigr)P_k(\AR_{p,n}^\pm)\Psi_{p,n}^\pm(0,0;z)  
\\
&= \left((\AR_{p,n}^\pm)^{\ell}z^p+\!\!\!\sum_{i+j=\ell -1}\!\!(\AR^\pm_{p,n})^i[z^p,\AR_{p,n}^\pm](\AR^\pm_{p,n})^j\right)\!\!P_k(\AR_{p,n}^\pm)\Psi_{p,n}^\pm(0,0;z)
\\
&= \left((\AR^{\pm}_{p,n})^{\ell}z^p-\ell(\AR^{\pm}_{p,n})^{\ell-1}\right)   P_k(\AR_{p,n}^\pm)\Psi_{p,n}^\pm(0,0;z)\in \WR^\pm_{p,n} ,
\end{aligned}$$
from (\ref{97}) and (\ref{105}) upon using (\ref{113}), yielding (\ref{107}); also the codimension statement is obvious as well as (\ref{107bis}). To see that $\Psi_{p,n}^\pm(x,0;z)$, as given by (\ref{kvpwavefunctions}), are indeed the wave functions at ${\mathbf t} = 0$, by precisely the argument of Lemma \ref{lemma3}, we need to show (see (\ref{19}))
$$\mathcal A^\pm_{p,n}(z)\Psi_{p,n}^\pm(x,0;z) = \pm D \Psi_{p,n}^\pm(x,0;z),$$
which follows from (\ref{kvpwavefunctions}), using (\ref{96}) and (\ref{111}), concluding the proof of the theorem.
\qed

\section{p--Airy kernels, vertex operators and Virasoro}

In this section we write the basic kernels of our theory in terms of KP wave operators, leading to Virasoro identities for the associated Fredholm determinants (\ref{Virasoro}) and various useful identities (\ref{43''}), (\ref{48}) and (\ref{3.48}) for the kernels, expressing them in terms of wave functions or double contours integrals.\\
Vertex operators typically generate Darboux transformations in integrable systems at the level of the tau functions. The KP vertex operator $\BX({\mathbf t},y,z)$ is defined as:
\be
\BX({\mathbf t},y,z):=\frac{1}{z-y}e^{\sum_1^{\iy}(z^i-y^i)t_i}e^{\sum_1^{\iy}(y^{-i}-z^{-i})\frac{1}{i}\frac{\pl}{\pl t_i}}.
\label{vertex}\ee
Given two integers $p,n \geq0 $ and $w\in\BR$ let us consider the KP wave functions at ${\mathbf t} = 0$
\bea
	\Psi_{p,n}^\pm(x,0;z) &:=& \dis\sqrt{\dis\frac{\pm p}{2\pi}}z^{\frac{p-1}{2}\mp n}{\rm e}^{\mp \frac{p}{p+1}z^{p+1}}\dis\int_{\Gamma_p^\pm}{\rm e}^{\mp \frac{y^{p+1}}{p+1} \pm xy}(y-w)^{\pm n}dy,\label{40'a}
\label{3wf}\eea
where the paths $\Gamma_p^{\pm}$ are chosen appropriately as described in the previous section.\\ 
According to sub--sections 2.2 (for the case $n=0$) and 2.3 (for the case $n >0$) the objects in (\ref{3wf}) are wave functions for the p--reduced KP hierarchy and for the 1--vector $p$--reduced KP hierarchy respectively. The corresponding (pseudo)--differential Lax operators, at ${\mathbf t} = 0,$ read
\be\begin{aligned}
\LR_{p,n}^\pm(x,0)&:= (\pm D)^p-x- n(\pm D-w)^{-1} 
\\&= 
(\pm D)^p-x-n{\rm e}^{\pm xw }(\pm D)^{-1}{\rm e}^{\mp xw}.
\end{aligned}\label{3lo}\ee 
In the sequel, when the subscript $n$ is omitted, we mean $n = 0$, coherently with the notation in the subsections 2.1 and 2.2.\\
We can now define the basic objects of our theory, remembering $D^{-1}$ means $\dis\int dx$. 
\begin{definition}\label{3.1}
	Given the wave functions (\ref{3wf}) we define the integral kernels $k^{(p)}_{x,{\mathbf t}},K^{(p)}_{x,{\mathbf t}}$ as
\bea
	 k^{(p)}_{x,{\mathbf t}}(z,z')&:=&D^{-1}\left(~\Psi^-_{p,n}(x,{\mathbf t};z)\Psi^+_{p,n}(x,{\mathbf t};z')\right),\\
K^{(p)}_{x,{\mathbf t}}(\lb,\lb')&:=& {\rm e}^{-\frac{p}{p+1}z^{p+1}}\left.
 \frac{k^p_{x,{\mathbf t}}(z,z')}{2\pi p z^{\frac{p-1}{2}+n}z'^{\frac{p-1}{2}-n}}
{\rm e}^{\frac{p}{p+1}z'^{p+1}}
\right|_{{z=\lb^{1/p}}\atop{z'=\lb'^{1/p}}}.
\label{defkernels}\eea
Also let $E := \bigcup^r_{i=1}[a_{2i-1},a_{2i}]\subset\BR^+$ be a disjoint union of intervals and let $$E^{\frac{1}p} := \{x \in \BR^+\;\mathrm{such}\,\mathrm{that}\; x^p\in E\}.$$ We define, for a tau function $\tau({\mathbf t})$, another function $\tau_E({\mathbf t})$:
\be
	\tau_E({\mathbf t}):= {\rm e}^{^{-\dis \mu\displaystyle{\int_{E^{{1}/{p}}}}dz~\BX (t;\om z,\om'z)}}\tau({\mathbf t}).
\label{deftauE}\ee
\end{definition} 
The function $\tau_E({\mathbf t})$ will be shown to be a tau function itself in Proposition \ref{prop4}.\\
The following propositions, due to Adler, Shiota and van Moerbeke in \cite{ASvM}, will be crucial. In the following we denote ($:\;:$ means normal ordering)
$$W_i^{(1)}=\frac{\pl}{\pl t_i}+(-i)t_{-i},\; W^{(2)}_{\ell}=\sum_{i+j=\ell}:W_i^{(1)}W_j^{(1)}:-(\ell+1)W_\ell^{(1)}$$ 
and $c_{p,j}^{(0)}=\delta_{1,j}\dis\frac{p^2-1}{12 p^2}$.
This first proposition is a consequence of Theorem 6.1 of \cite{ASvM}.
\begin{proposition}\label{prop4-}
The kernel $k^{(p)}_{x,{\mathbf t}}(z,z')$ satisfies the equation (see (2.4) for the notation $\bar{\mathbf t}$)
 \be
  	k_{x,{\mathbf t}}^{(p)}(z,z')=\frac{\BX(\bar{\mathbf t},z,z')\tau(\bar{\mathbf t})}{\tau(\bar{\mathbf t})}
  \label{43''}\ee 
  where we have an identity of formal expansions in Sato's theory.
  \end{proposition}
  
This second proposition explain how to express the Fredholm determinant of $K^{(p)}_{x,{\mathbf t}}$ in terms of the KP tau functions we described here and in the subsequent sections.
  
  \begin{proposition}\label{prop4}
Consider a disjoint union of intervals $E := \bigcup^r_{i=1}[a_{2i-1},a_{2i}]\subset\BR^+$ and the Fredholm determinant $\det ( \un -2\pi\mu K^{(p)}_{x,{\mathbf t}} \raisebox{1mm}{$\chi$}{}_{E} )$. Given $\omega\neq\omega'$ two $p$--roots of the unity, the following equality is satisfied
 
 \be 
	\det ( \un -2\pi\mu K^{(p)}_{x,{\mathbf t}} \raisebox{1mm}{$\chi$}{}_{E} )= \frac{\tau_E(\bar{\mathbf t})}{\tau(\bar{\mathbf t})}.
\label{44}\ee 
 Moreover  $\tau_E$, as $\tau$, is a $k$--vector p--reduced KP tau function
  ($k=0,1$ when $n=0,n>0$ respectively) and the following Virasoro constraints are satisfied:

\bea
	&&\left(\delta_{1,j}(1-\delta_{n,0}) w\frac{\pl}{\pl{w}}+\frac{1}{2p}W^{(2)}_{(j-1)p}+W^{(1)}_{jp+1}+\frac{2n-p+1}{2p}W_{(j-1)p}^{(1)}\right)
	\left\{\begin{array}{l}
		\tau({\mathbf t})\\
		\\
		\tau_E({\mathbf t})
	\end{array}\right\} =\nonumber\\
	&&\left\{\begin{array}{l}
		-c_{p,j}^{(n)}\tau({\mathbf t})\\
		\\
		\left(-c_{p,j}^{(n)}+\dis\sum_1^{2r}a^j_i\frac{\pl}{\pl a_i}\right)\tau_E({\mathbf t})
	\end{array}\right\}
\label{Virasoro}\eea
where $c_{p,j}^{(n)}$ are constants, given above for the case $n = 0$.
These constraints are valid, for $n=0$, $\forall j\geq 0$ while, for $n >0$, they hold just for $j=0,1$.
\end{proposition}
 
 \noindent\emph{Sketch of the }\proof\\
Formula (\ref{44}) is just Corollary 7.2.2 of \cite{ASvM}, modulo the conjugation term\\
$z^{-n}{\rm e}^{-\frac{p}{p+1}z^{p+1}}{\rm e}^{\frac{p}{p+1}z'^{p+1}} z'^{n}$ in $K^{(p)}_{x,{\mathbf t}} $, which has no effect on the Fredholm determinant. Now we have to prove the Virasoro constraints and the fact that $\tau_E$ is a tau function.\\
First  let us consider the $n=0$ case. In order to deduce (\ref{Virasoro}) from the Corollary 7.2.2 of \cite{ASvM}, the first fact is that 
 %
%
 by  the invariance conditions (\ref{16}), we find that the point in the Grassmannian $\WR^+_{p}$
 satisfies $z^{jp}\AR_p^+ \WR^+_p\subset \WR^+_p$, with
$$
 z^{jp}\AR_p^+=\frac{1}{p}z^{(j-1)p+1}\frac{\pl}{\pl z}+z^{jp+1}- 
\frac{p-1}{2p} z^{(j-1)p},\quad j\geq 0,
$$
and so by (\ref{12}), one has $(\PR^+_{z^{jp}\AR_p^+})_-=0$, and in particular
$$
0=\frac{1}{\Psi^+_{p}}\left( \frac{1}{p}M^+_p(L^+_p)^{(j-1)p+1}+(L^+_p)^{jp+1}-\frac{p-1}{2p}(L^+_p)^{(j-1)p} \right)_-\Psi^+_{p},
$$
which by the Adler-Shiota-van Moerbeke correspondence \cite{ASvM2} leads to \be\left(\frac{1}{2p}W^{(2)}_{(j-1)p}+W^{(1)}_{jp+1}-\frac{p-1}{2p}W^{(1)}_{(j-1)p}+c_{p,j}^{(0)}\right)\tau(t)=0,~~\mbox{for $j\geq 0$.} \label{47}\ee
(Here and in the sequel we denote with the subscripts ``$p$'' or ``$p,n$'' the Lax operators related to the particular solutions of the relevant KP hierarchy we are dealing with.)
The first term in the operator contains $\frac{1}{p}(p+1)t_{p+1}\pl/\pl t_{jp+1}$ and thus the shift $t_{p+1}\mapsto t_{p+1}-\frac{p}{ p+1}$ has the virtue to eliminate the $W^{(1)}_{jp+1}$ term in (\ref{47}), thus yielding
$$
\left(\frac{1}{2p}W^{(2)}_{(j-1)p}-\frac{p-1}{2p}W^{(1)}_{(j-1)p}+c_{p,j}^{(0)}\right)\tau=0.
$$
Both, Corollary 3.2.1 and Theorem 4.1 in \cite{ASvM}, and $\om^p=\om^{' p}=1$, yields
\be
\frac{\pl}{\pl z}(z^{(j-1)p +1}\BX({\mathbf t},\om z,\om'z))=\left[\frac{1}{2}W_{(j-1)p}^{(2)}-\frac{p-1}{2}W_{(j-1)p}^{(1)} +pc_{p,j}^{(0)}~,~\BX({\mathbf t},\om z,\om 'z)\right],\label{47'}
\ee
 without the $W_{(j-1)p}^{(1)}$-term. It is legitimate to add this term, because
 $$(v^\ell-u^\ell)\BX ({\mathbf t},u,v)=[W_\ell^{(1)},\BX({\mathbf t},u,v)],$$ and thus 
 $$[W_\ell^{(1)},\BX({\mathbf t},u,v)]=0\quad \mathrm{for}\; u=\om z,v=\om' z\quad\mathrm {and}\; p|\ell.$$ Then doing the shift again  $t_{p+1}\mapsto t_{p+1}+\frac{p}{p+1}$ reintroduces the $W_{jp+1}^{(1)}$-term again in (\ref{47'}). From the identity (\ref{47'}) and (\ref{47}), it then follows from the arguments of Theorem 4.1 of \cite{ASvM} that $\tau_E$ defined by (\ref{deftauE}) satisfies
 \be 
 \left(-\sum_1^{2r}a^j_i \frac{\pl}{\pl a_i} +
 \Bigl( \frac{1}{2p}W^{(2)}_{(j-1)p}+W^{(1)}_{jp+1}-\frac{ p-1}{2p}W_{(j-1)p}^{(1)}+c_{p,j}^{(0)}\Bigr)\right)\tau_E=0.
\label{46'} \ee
That $\tau_E$ is actually a $\tau$-function follows from several important facts:\\

({\em i}) if $\tau$ is a $\tau$-function, then $e^{a\BX}\tau=(1+a\BX)\tau$ is as well.\\

({\em ii}) the vertex operators for different indices commute:
$$ 
[\BX({\mathbf t},\lb,\mu),\BX({\mathbf t},u,v)]=0,\quad \mbox{for }u\neq\mu,\lb\neq v.
$$

({\em iii}) The integral in the exponential (\ref{deftauE}) is the limit of a Riemann sum, which using the higher Fay identities guarantees that $\tau_E$ is a $\tau$-function and it is expressible as a Fredholm determinant, as explained in \cite{ASvM}. Moreover, for $n=0$, $\tau$ and $\tau_E$ are $p$-reduced KP $\tau$-functions; indeed $\tau$ is, because the above $\WR^+_p$ satisfies $z^p\WR^+_p\subset \WR^+_p$. Also $\tau_E(t)$ is, because $\BX({\mathbf t};\om z,\om'z)$ is missing $t_{ip}$ and $\frac{\pl}{\pl t_{i p}}$, since $\omega^{ip}=\omega'^{ip}=1$ for $i\geq 1$, concluding the proof for the $n=0$ case.\\

Now for $k=1$ (i.e. $n > 0$), we have that the $1$--vector p--constrained tau function is also characterized by the fact that $\tau$ and $(\pl/\pl t_p)\tau$ are both tau functions \cite{vdLcKP}. Since the vertex operator in (\ref{44}) is free of $t_{ip}$, $(\pl/\pl t_{ip})$, $i\in\BZ_+$, (\ref{44}) yields
 $$
 \frac{\pl \tau_E}{\pl t_p} =e^{-\mu\dis\int_{E^{ {1}/{p}}}dz~\BX({\mathbf t},wz,w'z)} \frac{\pl \tau }{\pl t_p}
 $$
and this is a $\tau$-function since $(\pl/\pl t_p)\tau$ is a $\tau$-function and ${\rm e}^{-\mu\int_{E^{{1}/{p}}}dz~\BX({\mathbf t},wz,w'z)}$ takes $\tau$-functions to $\tau$-functions, as mentioned before. In particular, therefore $\tau_E$ is a $\tau$-function as well as $(\pl/\pl t_p)\tau_E$, and hence $\tau_E$ is a $1$--vector p--constrained $\tau$-function.\\
To see (\ref{Virasoro}) for the $n > 0$ case we follow roughly the $n=0$ argument. By (\ref{k-red}) (\ref{107}) and (\ref{107bis}) (we recall that here $k=1$), one has the following inclusions:
 $$
 \AR_{p,n}^+\WR^+_{p,n}\subset \WR^+_{p,n},\; z^p(\AR_{p,n}^+-w)\WR^+_{p,n}\subset \WR^+_{p,n}, ~\mbox{for}~
 \AR_{p,n}^+(z)=z+\frac{1}{pz^p} \left(z\frac{\pl}{\pl z}+n-\frac{p-1}{2}\right).
 $$
 By (\ref{12}), one has $(\PR^+_{\AR^+_{p,n}})_-=0$, and thus one has (\ref{47}) for $j=0$ with 
 $$-\frac{p-1}{2}\mapsto n-(\frac{p-1}{2}).$$ 
This proves, by the same argument as before, the identity (\ref{Virasoro}) for $j=0$.\\
For the case $j=1$, the relation follows from $(P^+_{(z^p\AR_{p,n}^+ - wz^p)})_-=0$, and so we have one extra term, not present in the $k=0$ case, due to the presence of $wz^p$, yielding (\ref{Virasoro}) with $k = n = 0$ for $j=1$, with one additional term, 
 \be
 \left(-\sum_1^{2r}a_i\frac{\pl}{\pl a_i}+\frac{1}{2p}W_0^{(2)}+W^{(1)}_{p+1}- {wW_p^{(1)}}+c_{p,1}^{(n)}\right)\tau =0.
 \ee
 It suffices to show $(\pl/\pl t_p)\tau = - (\pl/\pl w)\tau$ to conclude the proof of (\ref{Virasoro}). 
 To see this, notice that the integral $\Psi_{p,n}^+(x,0,z)$, as in (\ref{40'a}), readily satisfies
 $$
 \frac{\pl\Psi_{p,n}^+}{\pl w}(x,0;z)= -n(D-w)^{-1}\Psi_{p,n}^+(x,0;z)=(L^+_{p,n})^p_-\Psi_{p,n}^+(x,0;z)
 ,$$
 and upon differentiation by $x$,
 $$
 \frac{\pl}{\pl w} \Bigl(\frac{\pl^{ }}{\pl x }\Bigr)^\ell\Psi_{p,n}^+(x,0;z)  = (L^{+p}_{p,n})_- \Bigl(\frac{\pl^{ }}{\pl x }\Bigr)^\ell\Psi_{p,n}^+(x,0;z),
 $$
 and so, acting on $\WR^+_{p,n}$, as defined in (\ref{105}),
 \be
\left( \frac{\pl}{\pl w}-L^{+p}_{p,n}\right)_-=0.
 \label{neg}\ee
 Now let's recall the standard Sato formula (\ref{SatoFormula}), here rewritten as
 \be 
 	\Psi^\pm(x,{\mathbf t};z)={\rm e}^{\pm(xz+\sum_1^\iy t_i z^i)}\frac{{\rm e}^{\mp\eta} \tau(\bar{\mathbf t})}{\tau(\bar{\mathbf t})},~~\mbox{with}~ \eta:=\sum_1^\iy \frac{z^{-i}}{i}\frac{\pl}{\pl t_i}.
\label{tau3}\ee
The formula (\ref{neg}) leads to
$$
(e^{-\eta}-1)\left(\dis\frac{-\dis\frac{\pl\tau}{\pl w}}{\tau}\right)=\frac{-\dis\frac{\pl}{\pl w}\Psi_{p,n}^+(x,{\mathbf t};z)}{\Psi_{p,n}^+(x,{\mathbf t};z)}=-\frac{(L^{+p}_{p,n})_-\Psi_{p,n}^+(x,{\mathbf t};z)}{\Psi_{p,n}^+(x,{\mathbf t};z)}=(e^{-\eta}-1) \left(\frac{\dis\frac{\pl\tau}{\pl t_p}}{\tau}\right),
$$
using in the first equality straight differentiation and in the last equality again the Adler-Shiota-van Moerbeke correspondence \cite{ASvM2}, (remembering $\eta$ from (\ref{tau3})).
This shows 
$$\frac{\pl\tau}{\pl t_p}= -\frac{\pl\tau}{\pl w},$$ as claimed, concluding the proof of the Proposition \ref{prop4}. \qed

Proposition \ref{prop4} then leads to the following theorem involving bona fide analytic identities, rather than identities involving just formal series. In the following we denote with the subscript $t_{\geq p = 0}$ the locus given by $\left\{t_i = 0\;\;\forall i\geq p\right\}$. 

\begin{theorem}\label{theo:5}  Setting all $t_i=0,\;\forall i \geq p$, one has the following kernel identity for the kernel defined in (\ref{defkernels}),
\be
K_{x,{\mathbf t}}^{(p)} (z^p,z'^p)\Bigr|_{t_{\geq p}=0}= 
 \displaystyle{\frac{1}{(2\pi i)^2}\int_{\Gamma_p^+}du\int_{\Gamma^-_p}dv\frac{e^{-V_p(u)+(z'^p+x)u}}{e^{-V_p(v)
+(z^p+x)v}}\left(\frac{u-w}{v-w}\right)^{n}\frac{1}{u-v}},
\label{48}\ee
where $V_p(u)$ is a polynomial
\be
V_p(u):=\frac{u^{p+1}}{p+1}+\sum_{i = 0}^{p-2}\theta_i\frac{u^{i+1}}{i+1},\quad
\theta_i :=\theta_i(t_1,\ldots,t_{p-1}),
\label{49}\ee
with $\theta_i$ polynomials implicitly given in terms of $t_1,...,t_{p-1}$, by solving the equation $w=V_p'(u)$ for $u$ in terms of a series in large $w$, as in (i) and identifying it with another series, as in (ii); thus %
\bea
u&\stackrel{(i)}{=}&w^{\frac1p}-\frac{1}{p}\theta_{p-2}w^{-\frac1p}-\frac{1}{p}\theta_{p-3}w^{-\frac2p}-
\frac{1}{p}\left(-\frac{p-3}{2p}\theta^2_{p-2}+\theta_{p-4}
\right)w^{-\frac3p}\nonumber\\
& &\hspace*{5cm}+\dots +O(w^{-1-\frac1p})\label{3.18}\\
&\stackrel{(ii)}{=}&w^{\frac1p}+\frac{1}{p}\sum_1^{ p-1}(p-j)t_{p-j}w^{-\frac{j}{p}}+O(w^{-1-1/p})\nonumber
.\eea
\end{theorem}
\begin{example}
The first $\theta_i$ satisfy the following equations:
$$\mbox{$
\dis\frac{1}{p-1}\theta_{p-2}=- t_{p-1},~~~~\dis\frac{1}{p-2}\theta_{p-3}= -t_{p-2},~~~~\dis\frac{1}{p-3}\theta_{p-4}=- t_{p-3}+\dis\frac{1}{2p}(p-1)^2t^2_{p-1},\ldots
$}$$
\end{example}

The proof of Theorem \ref{theo:5} requires Proposition \ref{lemma:2}, which itself is based on the following Lemma:

\begin{lemma} \label{lemma:1} Consider the solution to the p--reduced KP--hierarchy coming from $\WR^+_p\subset\mathrm{Gr}$ as in Theorem \ref{theo1}. The associated differential Lax operator $\LR^+_p(x,\mathbf{t})=L^{+p}_p(x,\mathbf{t})$ satisfies
\be
\left[\LR^+_p,\left(1+\frac{p+1}{p}t_{p+1}\right)D+\frac{1}{p}\sum^{\iy}_{k=2+p\atop{k\neq ip}}kt_k(\LR^{+\frac{k-p}{p}}_p)_+\right]=1,
\label{51}\ee
and thus for $t_{\geq p}=0$, one has
\be
\left[\LR^+_p(x,t_1,...,t_{p-1}),D\right]=1.
\label{52}\ee
\end{lemma}
\noindent{\it Proof of Lemma \ref{lemma:1}:} 
\, For notational convenience let's denote 
\be
	S_p:=\frac{\tau(\bar{\mathbf t}-[D^{-1}])}{\tau(\bar{\mathbf t})}= W_p{\rm e}^{-\sum_1^{\iy}t_iD^i}
\label{Swaveoperator}\ee
with $W_p$ the dressing operator as in (\ref{waveoperator}) associated to this particular solution. Our first claim is that the corresponding operator $M_p^+$ of (\ref{7}) satisfies the equation
\be
M^+_p=S_p x S^{-1}_p+\sum_{k = 1}^{\iy}kt_k(L^+_p)^{k-1}.
\label{54}\ee
Indeed, from (\ref{7}), one has
\be
M^+_p=W_pxW^{-1}_p=S_p {\rm e}^{\sum_1^{\iy}t_kD^k} x {\rm e}^{-\sum_1^{\iy}t_kD^k}S^{-1}_p.
\label{55}\ee
 Since $[D,x]=1$, one has $[f(D),x]=f'(D)$, and thus
$$
\frac{\pl}{\pl t_i}{\rm e}^{\sum_1^{\iy}t_kD^k} x {\rm e}^{-\sum t_kD^k}={\rm e}^{\sum_1^{\iy}t_kD^k}[D^i,x]e^{-\sum_1^\iy t_kD^k}=iD^{i-1},
$$
and so 
\be
{\rm e}^{\sum_1^{\iy}t_kD^k}x{\rm e}^{-\sum_1^\iy t_kD^k}=x+\sum_{k = 1}^{\iy}kt_kD^{k-1}.
\label{56}\ee
Setting this formula into (\ref{55}) and using $S_pD^jS^{-1}_p=(L^+_p)^j$ yields (\ref{54}).\\
The conditions $z^p\WR^+_p \subset \WR^+_p$ and $\AR _p^+\WR^+_p\subset \WR^+_p$ implies, upon using (\ref{12}), that both $\LR^+_p = (L^+_p)^p$ and $\PR^+_{\AR_p^+}$ are differential operators. One then computes, using (\ref{10}), (\ref{12}) and  (\ref{54}) in the third line,
\be
\begin{aligned}
1&=[\AR_p^+,z^p]=[\PR^+_{z^p},\PR^+_{\AR_p^+}]
=[(\PR^+_{z^p}),(\PR^+_{\AR_p^+})_+]
\\
&=\left[\LR^+_p,\left(L^+_p + \frac1p {M^+_p(L^+_p)^{1-p}}-\frac{p-1}{2p}(L^+_p)^{-p}\right)_+\right]\\
&=\left[\LR^+_p,(S_pDS_p^{-1})_++\dis\frac1{p}{(S_px D^{1-p}S^{-1}_p)_+} +\dis\frac{1}{p}\sum^{\iy}_{k=1}kt_k\left((L^+_p)^{k-p}\right)_+-
 \dis\frac{p\!-\!1}{2p}
   (S_pD^{-p}S_p^{-1})_+ \right]\\
   &=\left[\LR^+_p,D+\dis\frac{1}{p}\sum^\iy_{k=p+1}kt_k\left(L^{+(k-p)}_p\right)_+\right]\\
&=\left[\LR^+_p, \left(1+ \dis\frac{p+1}{p} t_{p+1}\right)D+\frac{1}{p}\dis\sum^{\iy}_{k=p+2\atop{k\neq ip}}kt_k\left(L^{+ (k-p)}_p\right)_+\right]
\end{aligned}
\label{56'}\ee
yielding (\ref{51}) and hence  (\ref{52}), upon setting $t_{\geq p}=0$. \qed

\begin{proposition}\label{lemma:2}  

Consider the Lax operator $\LR_{p,n}^+(x,{\mathbf t})$ flowing off the initial condition $\LR_{p,n}^+(x,0)$ in (\ref{3lo}) together with its wave functions $\Psi^\pm_{p,n}(x,{\mathbf t};z)$ flowing off the initial conditions $\Psi_{p,n}^\pm(x,0;z)$ given in (\ref{3wf}). Then, evaluated at $t_{\geq p} = 0$, these quantities have the following form:
\be
\LR_{p,n}^+(x,t_1,...,t_{p-1})=V'_p(D)-x-n(D-w)^{-1},
\label{53}\ee
where $V_p(y)$ is defined in (\ref{49}), and
\be
\begin{aligned}
\Psi_{p,n}^\pm(x,t_1,...,t_{p-1};z)&={\rm e}^{\pm P({\mathbf t})}\dis\sqrt{\pm\dis\frac{p}{2\pi}}z^{\frac{p-1}{2}\mp n}{\rm e}^{\mp \frac{p}{p+1}z^{p+1}}\dis\int_{\Gamma_p^\pm}{\rm e}^{\mp V_p(y)\pm(x+z^p)y}(y-w)^{\pm n}dy,  
\end{aligned}
\label{53'}\ee
where the polynomial $P({\mathbf t}) = P(t_1,\ldots,t_{p-1})$ is determined by
$$
\frac{\pl}{\pl t_i}P({\mathbf t})=\left((V'_p(y))^{i/p}\right)_+\Big|_{y=0},\quad 1\leq i\leq p-1 \mbox{~and~}P(0)=0.
$$
\end{proposition}

\noindent{\it Proof of Proposition \ref{lemma:2}:} \, \\
We divide the proof in three steps; in the first step we prove (\ref{53}), for the case $n=0$, in the second step we prove (\ref{53'}) again for $n=0$ while in the third step we extend both formulas to the case $n>0$.\\
{\em Step 1}: Let $n=0$. Since $\LR^+_p(x,t_1,...,t_{p-1})$ is a differential operator of order $p$ satisfying the commutation relation (\ref{52}), namely $
\left[\LR^+_p(x,t_1,...,t_{p-1}),D\right]=1$, it must have the following form:
\be\label{57a}
\LR^+_p(x,t_1,...,t_{p-1})=(L_p^{+p})_+=D^p+\sum_{k=0}^{p-2}\theta_k(t_1,...,t_{p-1})D^k-x=:Q(D)-x,
\ee
and thus,\footnote{Since $(\LR^+_p)^{\frac{1}p}=D+\sum_{i=1}^{p-2}a_i(t)D^{-i}+a_p(t,x)D^{-p+1}+\ldots,(\LR_p^{\frac{n}p})_+$ is free of $x$ for $n<p$.}
\be
((\LR^+_p)^{{n}/{p}})_+= D^n+\sum^{n-2}_{i=0}c_{ni}(t_1,...,t_{p-1})D^i= 
 \Big(\left(Q(D)\right)^{{n}/{p}}\Big)_+,\quad 1\leq n\leq p-1,
\label{57'}\ee
%
with the $c_{ni}$ being polynomials in the $\theta_j$. One concludes that
\bean
\frac{\pl Q(D)}{\pl t_n}=\frac{\pl\LR^+_p}{\pl t_n}=[((\LR^+_p)^{{n}/{p}})_+,\LR^+_p]&=&[(Q(D)^{\frac{n}{p}})_+,Q(D)-x]\\
&=&[(Q(D)^{\frac{n}{p}})_+,-x]\\
&=&-\frac{\pl(Q(D)^{{n}/{p}})_+}{\pl D},
\eean
and thus, replacing $D$ by $y$,
\be
\frac{\pl Q(y)}{\pl t_n}=-\frac{\pl}{\pl y}(Q(y)^{\frac{n}{p}})_+.
\label{58}\ee
Next, for large y, solve the equation
\be
Q(y)=y^p+\sum_{k=0}^{p-2}\theta_k(t_1,...,t_{p-1})y^k
\ee
implicitly for $y=y(Q, t)$, namely
\bea
y&=&Q^{\frac1p}-\frac{1}{p}\theta_{p-2}Q^{-\frac1p}-\frac{1}{p}\theta_{p-3}Q^{-\frac2p}-
\frac{1}{p}\left(\theta_{p-4}-\frac{p\!-\!3}{2p}\theta^2_{p-2}
\right)Q^{-\frac3p}+\ldots+{\bf O}(Q^{-1-\frac{1}{p}}).\nonumber
\eea\be\label{60}\ee
Then, using (\ref{58}), one computes
\bean
0=\frac{d y}{d t_n}&=&\frac{d y}{d Q}\frac{\pl Q}{\pl t_n}+\frac{\pl y}{\pl t_n}
\\
&=&-\frac{\pl y}{\pl Q} \frac{\pl (Q(y)^{\frac{n}{p}})_+ }{\pl y}+\frac{\pl y}{\pl t_n} 
\\
&=&-\frac{\pl y}{\pl Q} \frac{\pl Q(y)^{\frac{n}{p}}}{\pl y} +\frac{\pl y}{\pl Q}{\bf O}\left(\frac{\pl}{\pl y}\left(\frac{1}{y}\right)\right)+\frac{\pl y}{\pl t_n}\\
&{=}&-\frac{\pl Q^{\frac{n}{p}}}{\pl Q}+ \frac{\pl y}{\pl t_n} +{\bf O}\left((Q^{\frac{1}{p}-1})\left(\frac{1}{Q^{2/p}}\right)\right)\\
&{=}&-\frac np Q^{\frac{n-p}p}+ \frac{\pl y}{\pl t_n} +{\bf O}\left( Q^{-\frac{1}{p}-1}  \right),
\eean
and thus
%
%
%
%
\be
\frac{\pl y}{\pl t_n}=\frac{n}{p}Q^{\frac{ n-p }{p}}+{\bf O}(Q^{-1-\frac{1}{p}}),\quad 1\leq n\leq p-1
;\ee
hence, $y$ as a function of $Q$ has the following form:
\be
y=Q^{\frac{1}{p}}+\frac{1}{p}\sum_{i=1}^{p-1}it_iQ^{ \frac{i-p}{p} }+{\bf O}(Q^{-1-\frac{1}{p}}).
\label{62}\ee
Equating (\ref{60}) and (\ref{62}), we solve for $\theta_i =\theta_i(t_1,...,t_{p-1})$, $1\leq i\leq p-1$ inductively as polynomials in the $t$, yielding $Q(y)=V'_p(y)$, where $V_p(y)$ is defined in (\ref{49}) and (\ref{3.18}) of Theorem \ref{theo:5}, concluding the proof of expression (\ref{53}) in Proposition \ref{lemma:2}.\\
{\em Step 2}: 
To show that $\Psi^\pm_p = \Psi^\pm_{p,0}$ defined in (\ref{53'}) are the KP wave functions, it suffices to show that they satisfy the first equations in (\ref{8}) and in (\ref{9'}). Let us first consider the former for $\Psi_{p}^+$. Indeed, the integral
\be
F_p(\lb):=\int_{\Gamma_p^+}{\rm e}^{-V_p(y)+\lb y}dy 
\label{63}\ee
satisfies
$$\begin{aligned}
0=\int_{\Gamma_p^+}\frac{\pl}{\pl y}{\rm e}^{-V_p(y)+\lb y}dy 
&=\int_{\Gamma_p^+}(-V'_p(y)+\lb){\rm e}^{-V_p(y)+\lb y}dy
  \\&=
 \left(-V_p'\left(\frac{\pl}{\pl\lb}\right)+\lb\right)F_p(\lb)
.\end{aligned}$$
Using the form (\ref{53}) of $\LR^+_p(x,t_1,...,t_{p-1})$ with $n=0$, it shows  that, setting $\lb =x+z^p$,  
\be
 \LR^+_p(x,t_1,...,t_{p-1})F_p(x+z^p)=(V_p'(D)-x)F_p(x+z^p)=z^pF_p(x+z^p),
\label{64}\ee
establishing $\LR^+_p(x,t_1,...,t_{p-1})\Psi_{p}^+=z^p\Psi^+_{p}$ for $\Psi^+_{p}$ as in 
(\ref{53'}) with $n=0$, and similarly for $\Psi_p^-$, establishing (\ref{8}).\\
We now establish (\ref{9'}); from (\ref{63}) and using (\ref{58}), namely 
$\frac{\pl}{\pl y}(  \frac{\pl}{\pl t_n} V_p(y)+(V_p'(y)^{n/p})_+)=0$,
integrated from 0 to $y$ and remembering $V_p(0)=0$, compute
\be
\begin{aligned}
\frac{\pl}{\pl t_n}F_p(x+z^p)&=-\dis\int_{\Gamma_p^+} \left(\frac{\pl}{\pl  
t_n}V_p(y)\right){\rm e}^{-V_p(y)+(x+z^p) y}dy
\\
&=\dis\int_{\Gamma_p^+}\left( (V'_p(y)^{\frac{n}{p}})_+ -(V'_p(y)^{\frac{n}{p}})_+ 
\Bigr|_{y=0}\right) {\rm e}^{-V_p(y)+(x+z^p) y}dy
\\
 &= (V'_p(D)^{\frac{n}{p}})_+ F_p(x+z^p)-(V'_p(y)^{\frac{n}{p}})_+\Bigr|_{y=0} 
  F_p(x+z^p).
 \end{aligned}
\label{64'}\ee
%
The expressions 
\be b_i(t):=
(V'_p(y)^{\frac{i}{p}})_+\Big 
\vert_{y=0}
 ,\quad 1\leq i\leq p-1,
\label{64''}\ee
are polynomials in $t_1,\ldots,t_{p-1}$, as follows from the proof of Lemma \ref{lemma:1}.  Observe by the KP flow compatibility conditions and by (\ref{57'}) that for $1\leq i,j\leq p-1$,
\bean
\frac{\pl b_i}{\pl t_j}-\frac{\pl b_j}{\pl t_i}&=&\left(\frac{\pl } 
{\pl t_j}(V'_p(D)^{\frac{i}{p}})_+ -
\frac{\pl }{\pl t_i}(V'_p(D)^{\frac{j}{p}})_+\right)\big\vert_{D=0}\\
& =&  \left(\frac{\pl }{\pl t_j}(\LR_{p}^{\frac{i}{p}})_+ -
\frac{\pl }{\pl t_i}(\LR_p^{\frac{j}{p}})_+\right)\big\vert_{D=0},\\
&=&\left[(\LR_p^{\frac{j}{p}})_+,(\LR_p^{\frac{i}{p}})_+\right]\big 
\vert_{D=0}\\
&=&\left[(V'_p(D)^{\frac{j}{p}})_+,(V'_p(D)^{\frac{i}{p}})_+\right]\big 
\vert_{D=0}= 0.
\eean
Thus, the polynomials $b_i(t)$ can then be represented as $b_i(t)=\frac{\pl}{\pl t_i}P(t) $ for some uniquely defined polynomial $P(t)$, modulo an  
additive constant, which we set to be 0, i.e., $P(0)=0$; so, we have that expression (\ref{64''}) equals
\be
 b_i(t)=(V'_p(y)^{\frac{i}{p}})_+\Big 
\vert_{y=0} = \frac{\pl}{\pl t_i} P(t)
,\quad 1\leq i\leq p-1.\ee
Setting this expression into (\ref{64'}), and defining $\Psi_{p}^+(x,t_1,...,t_{p-1};z)$ as the right hand side of the first expression (\ref{53'}), leads to the following differential equations satisfied by $\Psi_{p}^+$ for $1\leq n\leq p-1$:
\be
\begin{aligned}
\frac{\pl}{\pl t_n}\Psi_{p}^+(x,t_1,...,t_{p-1};z)&=
\sqrt{\frac{p}{2\pi}}z^{\frac{p-1}{2}}{\rm e}^{-\frac{p}{p+1}z^{p+1}}\frac{\pl}{\pl t_n}({\rm e}^{P(t)} F_p(x+z^p))
\\
&=\sqrt{\frac{p}{2\pi}}z^{\frac{p-1}{2}}{\rm e}^{-\frac{p}{p+1}z^{p+1}}(V_p'(D)^{\frac{n}{p}})_{_+} 
(e^{P(t)}  F_p(x+z^p))
\\
&=
\sqrt{\frac{p}{2\pi}}z^{\frac{p-1}{2}}{\rm e}^{-\frac{p}{p+1}z^{p+1}}(\LR_p^{+\frac{n}{p}})_{_+}({\rm e}^{P(t)}  F_p(x+z^p))
,~~\mbox{using (\ref{57'})}
\\
&=(\LR_p^{+\frac{n}{p}})_{_+}
\Psi_{p}^+(x,t_1,\ldots,t_{p-1};z)
\end{aligned}
\label{65} \ee
with the initial condition, by (\ref{40'a}) with $n=0$,  
$$
\Psi_{p}^+(x,0;z) = \sqrt{\frac{p}{2\pi}}z^{\frac{p-1}{2}}e^{-\frac{p}{p+1}z^{p+1}}e^{P(t)} F_p(x+z^p)\vert_{t=0}\;;
$$
thus establishing (\ref{9'}) for $\Psi_p^+$, and similarly for $\Psi_p^-$, and thus we find the identity (\ref{53'}) for $n = 0$ and conclude the proof of Proposition \ref{lemma:2} for $n=0$.\\
{\em Step 3}: 
The proof for the case $n>0$ follows from that of $n=0$ with some important modifications. Indeed, since $\PR^+_{z^p}=L^p_{p,n}\neq (L^p_{p,n})_+$, we find as in Lemma \ref{lemma:1} that
 \be
 [L^p_{p,n}(t_1,...,t_{p-1}),D]=1,
 \label{129}\ee
 but with
 \be
 L^p_{p,n}(t_1,...,t_{p-1})=V'_p(D)-x-n(D-w)^{-1},
 \label{130}\ee
 with $V_p(y)$ specified in Theorem \ref{theo:5}. Equations (\ref{129}) and (\ref{130}) are proven exactly as before. Indeed, one sets, analogous to the $n=0$ case, with $Q$ as in (\ref{57a}),
 $$
 L^p_{p,n}=Q(D)-x+(L^p_{p,n})_-,\mbox{~and so,~}(L^j_{p,n})_+=(Q(D)^{\frac{j}{p}})_+,\quad 1\leq j\leq p-1,
 $$
 and the same proof as before yields $Q(D)=V'_p(D)$. Indeed, one observes that $[(L^p_{p,n})_-, (Q(D)^{ {j}/{p}})_+]=0$, $1\leq j\leq p-1$, since $(Q(D)^{ {j}/{p}})_+$ is $x$-independent; thus we have: 
 $$
 \begin{aligned}
  -\frac{\pl (L^p_{p,n})_-}{\pl t_j}  &=\Bigl[L^p_{p,n},(Q(D)^{\frac{j}{p}})_+\Bigr]_-
\\
&= \Bigl[(L^p_{p,n})_+,(Q(D)^{\frac{j}{p}})_+\Bigr]_-+\Bigl[(L^p_{p,n})_-,(Q(D)^{\frac{j}{p}})_+\Bigr]_-=0,
\end{aligned}
 $$
 showing that $ {\pl (L^p_{p,n})_-}/{\pl t_j}  =0$; so $(L^p_{p,n})_-$ is unmoved by the first $p-1$ flows (but moved by the $t_p$ flow), yielding (\ref{58}) as before, etc. and so formula (\ref{130}) is indeed proven exactly as in the case $n=0$, while (\ref{53'}) is proven exactly as for the case $n=0$, completing the proof of Proposition \ref{lemma:2}.\qed
Now we can prove Theorem \ref{theo:5}.\\
\noindent{\it Proof of Theorem \ref{theo:5}:} \,First consider the case $n=0$. Setting $\Lambda(z,z'):=2\pi  p~z^{\frac{p-1} 
{2}}z'^{\frac{p-1}{2}}{\rm e}^{\frac{p}{p+1}z^{p+1}}{\rm e}^{-\frac{p}{p+1}z'^{p 
+1}}$ and using (\ref{43''}) together with the definitions in (\ref{defkernels}), 
\be
D\left(\Lambda(z,z')K_{x,{\mathbf t}}^{(p)} (z^p,z'^p)
 \right)=\Psi^-_{p}(x,{\mathbf t};z)\Psi^+_{p}(x,{\mathbf t};z').
\label{66'}
\ee
Then using $\frac{d}{dx}{\rm e}^{x(u-v)}=(u-v){\rm e}^{x(u-v)}$, the representation (\ref{53'}) for $\Psi^\pm_p = \Psi^\pm_{p,0}$ and (\ref{66'}) above, one checks
$$
\begin{aligned}
\lefteqn{ \hspace*{-1cm} D{\frac{\Lambda(z,z')}{(2\pi i)^2}\int_{\Gamma_p^+}du\int_{\Gamma^-_p}dv\frac{{\rm e}^{-V_p(u)+(z'^p+x)u}}{{\rm e}^{-V_p(v)
+(z^p+x)v}} \frac{1}{u-v}}}\\
&=\frac{\Lambda(z,z')}{(2\pi i)^2} 
 \int_{\Gamma^-_p}{\rm e}^{V_p(v)-(x+z^p)v}dv
 \int_{\Gamma_p^+}{\rm e}^{-V_p(u)+(x+z'^p)u}du 
\\
&=\Psi^-_{p}(x,{\mathbf t};z)\Psi^+_{p}(x,{\mathbf t};z') \Bigr|_{t_{\geq p}=0},~~\mbox{using (\ref{53'})},\\
&=
D \Lambda(z,z')K_{x,{\mathbf t}}^{(p)} (z^p,z'^p)\Bigr|_{t_{\geq p}=0}
 ,~~\mbox{using (\ref{66'})},
\end{aligned}
$$
thus yielding the desired identity (\ref{48}) of Theorem \ref{theo:5}, except for the differentiation $D$. To do the identification, without the $D$, one hits these identities with $D^{-1}$, sets $D^{-1}{\rm e}^{x(u-v)}=\frac{{\rm e}^{x(u-v)}}{u-v}$ and then identifies the formal expansions (of the Sato theory) of the wave functions $\Psi^\pm_{p,0}$ of Proposition \ref{prop4} (which can ultimately be traced to Lemma \ref{lemma3}) with the asymptotic expansions of the double integral (which can be ultimately be traced to Theorem \ref{3'}), which is an analytic object, concluding the proof of  Theorem \ref{theo:5} for $n=0$. This proof can be adapted without trouble to $n>0$.\qed
 
\example For $p=2,3$ we find from Theorem \ref{theo:5} that

$$\begin{aligned}
V_2(u)&= \frac{u^3}{3}-t_1u   
\\
V_3(u)&= \frac{u^4}{4}-t_2u^2-t_1u.
\end{aligned}$$
Hence, upon choosing the appropriate contours $\Gamma_p^\pm$ as in Corollary (\ref{contornigiusti}), we recognize, for $n = 0$, the well known expressions of of the Airy and the Pearcey kernel as double contour integrals. The case $n > 0$ corresponds, for $p = 2$, to the case of the kernel for the Airy process with outliers, studied in \cite{ADvM} and \cite{ADvMVa}. For $p=3$ we recover the kernel for the Pearcey process with inliers, again studied in \cite{ADvMVa}.
\begin{lemma}\label{lemma8} For $n=0$ the $\tau$-functions corresponding to the points in the Grassmannian $\WR^+$ such that:
$$
z^p\WR^+_p\subset \WR^+_p,\quad \AR_p^+(z)\WR^+_p\subset \WR^+_p,
$$
evaluated on the locus $t_i=0$, $i\geq p$, i.e. the so-called topological tau function $\tau^{(p)}_0(t_1,t_2,...,t_{p-1})$, are completely determined alternatively by\footnote{The residue is evaluated at $u = \infty$, i.e. it is the coefficient of $u^{-1}$, for $u^{-1}$ small.}
\be
\frac{\pl}{\pl t_1}\frac{\pl}{\pl t_i}\log\tau_0^{(p)}={\rm res}_u(V'_p(u))^{\frac{i}{p}},\quad 1\leq i\leq p-1,
\label{71}\ee
and the KP hierarchy equations containing just $\pl_1,\pl_2,\ldots,\pl_p$, or equivalently

\be
\frac{\pl}{\pl t_i}\log\tau_0^{(p)}=-\frac{p}{p+i}{\rm res}_u(V'_p(u))^{\frac{p+i}{p}},\quad 1\leq i\leq p-1,
\label{72}\ee
with $V_p(u)$ given in Theorem \ref{theo:5}. So in particular we find:
\be
\begin{aligned} 
\log\tau_0^{(2)}(t_1)&= -\dis\frac{1}{12}t_1^3,
\\
\log\tau_0^{(3)}(t_1,t_2)&= -\dis\frac{1}{3}t_1^2t_2-\dis\frac{2}{27}t^4_2,
\\
\log\tau_0^{(4)}(t_1,t_2,t_3)&= -\dis\frac{3}{8}t_1^2t_3-\dis\frac{1}{2}t_1t_2^2-\dis\frac{9}{16}t^2_2t^2_3-\dis\frac{81}{1280}t^3_3.
\end{aligned}
\label{73}\ee
\end{lemma}
We now give another representation of the kernel in terms of the wave and dual wave functions:
\begin{theorem}\label{theorem3.6} 
   Setting all $t_i=0, i \geq p$, one has the following  kernel identities for the kernel $K_{x,{\mathbf t}}^{(p)}$ defined in (\ref{defkernels}):
\be
\begin{array}{lll} 
\lefteqn{ K_{x,{\mathbf t}}^{(p)} (z^p,z'^p)\Bigr|_{t_{\geq p}=0}=\dis\frac{\varphi(z,z')}{z'^{p}-z^p} 
 \left(\dis { \sum_{k+\ell =p-1}D^k\Psi^+_{p,n}(z')(-D)^{\ell}\Psi^-_{p,n}(z)}\right.+}\\ \\
\left.\dis{\sum_{i=1}^{p-2}\theta_i\!\!\!\!\!\sum_{k+\ell =i-1}}\!\!\!\!D^k\Psi^+_{p,n}(z')(-D)^{\ell}\Psi^-_{p,n} (z)+n(D-w)^{-1}\Psi_{p,n}^+(z')(-D-w)^{-1}\Psi^-_{p,n}(z)\!\!\right)\!\!_{\Big| t_{\geq p}=0}
\end{array}
\label{48'}\ee
The prefactor in (\ref{48'}) reads $$\varphi(z,z'):= \dis\frac{i {\rm e}^{\frac{p}{p+1}(z'^{p+1}-z^{p+1})}}{2\pi p (z z')^{\frac{p-1}2}}\left(\frac{z'}{z}\right)^n$$ and $\Psi^\pm_{p,n}(z) =  \Psi^\pm_{p,n}(x,\mathbf{t},z)_{| t_{\geq p}=0}$ are given explicitly in Proposition \ref{lemma:2} (the dependence of $\Psi^\pm_{p,n}$ on the variables $(x,{\mathbf t})$ has been suppressed for notational's convenience).
\end{theorem}
Theorem \ref{theorem3.6} is an immediate consequence of (\ref{53'}) and the following useful lemma.
\begin{lemma}\label{lemma:3}
Let us denote
$$
\psi_p^\pm(z^{p})=\frac{1}{2\pi i}\int_{\Gamma_p^\pm}{\rm e}^{\mp V_p (u)\pm (x+z^{p})u}(u-w)^{\pm n}du.
$$
Then we have
\bea
&&K^{(p)}_{x,{\mathbf t}}(z^p,z^{'p})= \frac{1}{z'^{p}-z^p}\left(\dis{\sum_{k+\ell =p-1}}(D^k\psi_p^+(z'^p))(-D)^{\ell}\psi^-_p(z^p) \right.\label{3.48}\\
&+&\left. 
\sum_{i=1}^{p-2}\theta_i\dis{\sum_{k+\ell =i-1}} (D^k\psi^+_p(z'^p))(-D)^{\ell}\psi^-_p(z^p)+n(D-w)^{-1}\psi_p^+(z')(-D-w)^{-1}\psi^-_p(z) \right)\nonumber
\eea
\end{lemma}
 \noindent{\it Proof of Lemma \ref{lemma:3}}: \, Observe that, because of our choice of $\Gamma_p^\pm$,
 \bean
 0&=&\frac{1}{(2\pi i)^2}\int_{\Gamma_p^+}du\int_{\Gamma_p^-}dv\left(\frac{\pl}{\pl u}+\frac{\pl}{\pl v}\right)\left(\frac{u-w}{v-w}\right)^n
\left[
 \frac{{\rm e}^{-V_p (u)+(x+z'^{p})u}}{{\rm e}^{-V_p (v)+(x+z^{p})v}}\frac{1}{u-v}\right]\\
 \\
 &=&\frac{1}{(2\pi i)^2}\int_{\Gamma_p^+}du\int_{\Gamma_p^-}dv\\
 &&\frac{{\rm e}^{-V_p (u)+(x+z'^{p})u}}{{\rm e}^{-V_p (v)+(x+z^{p})v}}
  \left(\frac{u-w}{v-w}\right)^n\left[\left(\frac{-V_p' (u)+V'_p(v)+z'^{p}-z^p}{u-v}\right)-\frac{n}{(u-w)(v-w)}\right]\\
 \\
 &=&- \frac{1}{(2\pi i)^2}\int_{\Gamma_p^+}du\int_{\Gamma_p^-}dv
  \frac{{\rm e}^{-V_p (u)+(x+z'^{p})u}}{{\rm e}^{-V_p (v)+(x+z^{p})v}}\left(\frac{u-w}{v-w}\right)^n\left[\left(\frac{V'_p(u)-V'_p(v)}{u-v}\right)+\frac{n}{(u-w)(v-w)}\right]\\
 \\
 &+&\frac{z'^p-z^p}{(2\pi i)^2}\int_{\Gamma_p^+}du\int_{\Gamma_p^-}dv
 \frac{{\rm e}^{-V_p (u)+(x+z'^{p})u}}{{\rm e}^{-V_p (v)+(x+z^{p})v}}\left(\frac{u-w}{v-w}\right)^n\frac{1}{u-v}.
 \eean
The proof is finished upon using (\ref{48}) in the last expression and noticing that, by (\ref{49}),
\bean
\lefteqn{\frac{1}{(2\pi i)^2} \int_{\Gamma_p^+}\!\!\!\!\!du\int_{\Gamma_p^-}\!\!\!\!\!dv  \frac{{\rm e}^{-V_p (u)+(x+z'^{p})u}}{{\rm e}^{-V_p (v)+(x+z^{p})v}}\left(\frac{u-w}{v-w}\right)^n \left[\left(\frac{V'_p(u)-V'_p(v)}{u-v}\right)+\frac{n}{(u-w)(v-w)}\right]}\\
\\
&=&\frac{1}{(2\pi i)^2} \int_{\Gamma_p^+}\!\!\!\!\!du\int_{\Gamma_p^-}\!\!\!\!\!dv  \frac{{\rm e}^{-V_p (u)+(x+z'^{p})u}}{{\rm e}^{-V_p (v)+(x+z^{p})v}}\left(\frac{u-w}{v-w}\right)^n\!\left(\frac{u^p-v^p}{u-v}+\sum_{i=1}^{p-2}\theta_i\frac{ u^i-v^i }{u-v}+\frac{n}{(u-w)(v-w)}\right)\\
\\
&=&\frac{1}{2\pi i} \int_{\Gamma_p^+}\!\!\!\!\!du~{\rm e}^{-V_p (u)+(x+z'^{p})u}(u-w)^n\frac{1}{2\pi i} \int_{\Gamma_p^-}\!\!\!\!\!dv \frac{~{\rm e}^{V_p(v)-(x+z^{p})v}}{(v-w)^n}
\\
&&\hspace{5cm} \left(\sum_{k+\ell =p-1}u^kv^{\ell}+\sum^{p-2}_{i=1}\theta_i\sum_{k+\ell =i-1}u^kv^{\ell}+\frac{n}{(u-w)(v-w)}\right)\\
\\
&=&\!\!\!\sum_{k+\ell =p-1}(D^k\psi_p^+)(-D)^\ell\psi^-_p+\sum^{p-2}_{i=1}\theta_i\sum_{k+\ell =i-1}(D^k\psi_p^+)(-D)^{\ell}\psi_p^-+n(D-w)^{-1}\psi_p^+(-D-w)^{-1}\psi_p^-.
\eean\qed

\example Let us set $n=0$;  for $p=2,3$, one uses again the contours chosen in Corollary \ref{contornigiusti}; we recover using Lemma \ref{lemma:3} the expressions of the Airy and the Pearcey kernel as integrable kernels \`a la Its--Izergin--Korepin--Slavnov \cite{IIKS}.  Indeed, for $p=2$, denoting with $\Ai(x)$ the usual Airy function, we find, setting $Df = f'$, that
$$K^{(2)}_{x,t_1}(\lambda,\lambda')=\frac{\Ai'(x+t_1+\lambda')\Ai(x+t_1+\lambda)-\Ai(x+t_1+\lambda')\Ai'(x+t_1+\lambda)}{\lambda'-\lambda}$$
which is precisely the Airy kernel for $x+t_1=0$.
For $p=3$ let us define
\bean
	\psi_3^\pm(x,y)&:=&\frac{1}{2\pi i}\int_{\Gamma_3^\pm}{\rm e}^{\mp \frac{u^4}4\pm yu^2\pm xu}du;
\eean
which is nothing but the so--called Pearcey function together with its adjoint.
We obtain
\bean
	&&K^{(3)}_{x,t_1,t_2}(\lambda,\lambda')=\\
	&&\\
	&&\Bigg({\psi_3^+}''(x+t_1+\lambda',t_2)\psi_3^-(x+t_1+\lambda,t_2)-{\psi_3^+}'(x+t_1+\lambda',t_2){{}\psi_3^-}'(x+t_1+\lambda,t_2)\\
	&&\\
	&+&\psi_3^+(x+t_1+\lambda',t_2){{}\psi_3^-}''(x+t_1+\lambda,t_2)-2t_2\psi_3^+(x+t_1+\lambda',t_2)\psi_3^-(x+t_1+\lambda,t_2)\Bigg)(\lambda'-\lambda)^{-1},
\eean
the usual Pearcey kernel with parameter $t = t_2$, for $x + t_1 = 0$.

\section{PDEs for random matrix kernels}

The machinery of the previous sections will be used to compute PDEs for Fredholm determinants of kernels occurring in random matrix theory and Dyson Brownian motion, by specializing the kernels in (\ref{defkernels}), thus yielding a proof of Theorem \ref{introtheorem}. The PDEs will come from PDEs of the KP hierarchy exploiting the formula (\ref{44}). Then, using the Virasoro relations (\ref{Virasoro}), we will replace some of the partial derivatives in $t_i$ with partial derivatives in $a_i$, i.e. the end points of the disjoint union of intervals $E := \bigcup^r_{i=1}[a_{2i-1},a_{2i}]\subset\BR$. The next lemma provides the PDEs of the KP hierarchy given in the Hirota form which we shall use to generate the promised PDEs for the Fredholm determinants; we will denote with $p_i$ the Schur polynomials already defined in (\ref{Schur}). The following lemma can be deduced from arguments in \cite{DJKM1}.
\begin{lemma}\label{lemma6}The bilinear identity for KP (\ref{BHE}) generates two strings of Hirota relations,\footnote{We recall the standard notation for the Hirota symbol of two functions $f$ and $g$, associated with any polynomial of many variables
$$p(\pl_1,\pl_2,\ldots)f \circ g := p\left(\frac{\pl}{\pl t_1},\frac{\pl}{\pl t_2},\ldots\right)f(t_1+y_1,t_2+y_2,\ldots)g(t_1-y_1,t_2-y_2,\ldots)\big\vert_{\{y_i\}=0}$$}
\be
\begin{array}{lll} 
0&=&\dis\oint_{\iy}\dis\frac{dz}{2\pi i}\tau({\mathbf t}-[z^{-1}])\tau({\mathbf t'}+[z^{-1}])e^{\sum_1^{\iy}z^i(t_i-t'_i)}\big\vert_{t\mapsto t+\frac{1}{2}y\atop{t'\mapsto t-\frac{1}{2}y}}\\
\\
&=&\sum_{j=0}^{\iy}p_j(y)p_{j+1}(\pl_{\mathbf t})e^{-\frac{1}{2}\sum^{\iy}_1y_{\ell}\pl_{\ell}}\tau\circ\tau\\
\\
&=&\sum_{\ell =1}^{\iy}y_{\ell}\left(p_{\ell +1}(\pl_{\mathbf t})-\frac{1}{2}{\pl_1\pl_{\ell}}\right)\tau\circ\tau\\
\\
& &+\sum_{\ell =2}^{\iy}y_1y_{\ell -1}\left(p_{\ell +1}(\pl_{\mathbf t})-\frac{1}{4}
\pl_2\pl_{\ell -1}-\frac{1}{2}\pl_1p_{\ell}(\pl_{\mathbf t})\right)   \tau\circ\tau +\mathbf{O}(y_i^3),
\end{array}
\label{67}\ee
which are independent, for $\ell\geq 5$. The first string, denoted symbolically by $\BY_{\ell}$, is the standard KP hierarchy and we will denote twice the second one minus twice the first one by $\BY_{1,\ell -1}$
\be
\BY_{\ell}:\Bigl(p_{\ell +1}(\pl_{\mathbf t})-\frac{1}{2}\pl_1\pl_{\ell}\Bigr)\tau\circ\tau=0,  ~~~\BY_{1,\ell -1}:\Bigl(\pl_1\pl_{\ell}-\frac{1}{2}\pl_2\pl_{\ell -1}-\pl_1p_{\ell}(\pl_{\mathbf t})\Bigr)\tau\circ\tau=0.
\label{68}\ee

\end{lemma}
We are ready to state the main theorem of this section; we remember that, given a disjoint union of interval $E := \bigcup^r_{i=1}[a_{2i-1},a_{2i}]\subset\BR$, we defined in the introduction the two differential operators
$$\partial := \sum_i \frac{\pl}{\pl a_i}, \quad\quad \vr := \sum_i a_i\frac{\pl}{\pl a_i}.$$
Also we will denote 
\be
\BQ=\BQ_p(t_2,...,t_{p-1};E):=\log\det(I-K^{(p)}_{x,{\mathbf t}}\raisebox{1mm}{$\chi$}{}_{E} )\big\vert_{x=t_1=t_{\geq p} = 0}.
\label{75}\ee
\begin{theorem}\label{theorem9}
Each of the Hirota equations
\be
\BY_3,\ldots,~\BY_{p+1},~\BY_{1,4},\ldots,~\BY_{1,p}~~\mbox{and}~~2(p+2)\BY_{p+2}+(p+1)\BY_{1,p+1}, 
\label{75'}
\ee
gives rise to a non-linear PDE for the Fredholm determinant $\BQ$.
These PDE's only involve the differentials $\pl$ and $\vr$ with regard to the boundary points of $E$ and the $t$-partials $\pl_2,\pl_3,...,\pl_{p-1}$. 
\end{theorem}
Examples of PDEs are given below; also note that linear combinations of these equations can lead to new interesting equations.
\begin{corollary}\label{corolequations}
  $\BQ$ satisfies the following PDEs:\\

\noindent\boxed{\bf Case\;1\, (n=0):}\\

\noindent $\bullet$
{For all $p\geq 2$, $\BQ$ satisfies\\
\underline{the $\BY_3$-equation}: }
\be
\begin{aligned}
\pl^4\BQ+&6(\pl^2\BQ)^2 +\dt_{2,p}(2-4\vr)\pl\BQ
\\
&+(1-\dt_{2,p})\left(3\pl^2_2\BQ-4\left[3 \frac{ p-1}{p} t_{p-1}\pl+(1-\dt_{3,p})\pl_3\right]\pl\BQ\right)=0
\end{aligned} \label{76}\ee

\noindent$\bullet$ For all $p\geq 3$, $\BQ$ satisfies the equations (\ref{76}) and \\
\underline{the $\BY_4$-equation}:

\be
\begin{aligned}\pl_2\pl^3 \BQ+6(\pl_2\pl\BQ)\left(\pl^2\BQ-\mbox{$\dis\frac{1}{p}$}(p-1)t_{p-1}\right)
+\dt_{3,p}\Bigl((1-3\vr)\pl\BQ+2t_2\pl\pl_2\BQ\Bigr)
\\
+(1-\dt_{3,p})\left(2\pl_2\pl_3\BQ-3\left[\mbox{$\dis\frac{4}{p}$}(p-2)t_{p-2}\pl+(1-\dt_{4,p})\pl_4\right]\pl\BQ\right)=0,
\end{aligned}
\label{77} \ee
\underline{the $\pl_2\BY_3-\pl\BY_4$-equation}
\be\begin{aligned}
&\pl_2^3\BQ +2\left\{\pl_2\pl\BQ,\pl^2\BQ\right\}_{\pl}-2(1-\dt_{3,p})\pl_2\pl_3\pl\BQ 
+\dt_{3,p}(\varepsilon-2t_2\pl_2-2)\pl^2\BQ \\
&-\frac2p 
(1-\dt_{3,p})\left(( {p-1} )t_{p-1}\pl_2-2( {p-2} )t_{p-2}\pl-\frac p2 (1-\dt_{4,p})\pl\right) \pl^2\BQ=0,
\end{aligned}\label{78'}
\ee
\underline{the $ \pl^2\BY_3$-equation} (Boussinesq form of $\BY_3$ equation)
\be
 \pl^4 U+3\pl_2^2 U+6\pl^2 U^2-4(1-\dt_{3,p})\pl_3\pl U=0,\quad U:=\pl^2\BQ-\dis\frac{p-1}pt_{p-1}\label{Bous}.
\ee
 
 \bigbreak
 
 \noindent$\bullet$ For all $p\geq 4$, $\BQ=\BQ_p$ satisfies all previous equations
  and\\
\underline{the $\BY_5-\BY_{1,4}$-equation}:
\be\begin{aligned}
&\pl_2^2\pl^2 \BQ+\frac{2}{3}\pl_3\pl^3\BQ
+\frac{4}{3}\pl_3^2\BQ+4(\pl^2\BQ)(\pl_3\pl\BQ)+4(\pl_2\pl\BQ)^2
\\
&+2\left(\pl^2_2\BQ-\frac{p}{p+2}\pl_2{\rm res}_u(V'_p(u))^{\frac{p+2}{2}}\right)(\pl^2\BQ)-\frac{4}{p}(p-1)\pl_3\pl\BQ
  -12\mbox{$\dis\frac{p-3}{p}$}t_{p-3}(1-\dt_{4,p})\pl^2\BQ 
\\
&     -\mbox{$\dis\frac{16}{p}$}(p-2)t_{p-2}\pl_2\pl\BQ+\dt_{4,p}\left[1-4\vr +2t_2\pl_2+3t_3\pl_3\right]\pl\BQ+4(1-\dt_{4,p})(1-\dt_{5,p})\pl_5\pl\BQ=0
     \end{aligned}
     \ee
\noindent\boxed{\bf Case\; 2\,(n>0):}
\begin{flalign}
	&\underline{p=2}:\quad \pl^4\BQ+6(\pl^2\BQ)^2+(2-4(\vr-w\pl_w))\pl\BQ+3\pl^2_w\BQ=0&
 \end{flalign}
\begin{flalign}
 &\underline{p=3}:\quad  \pl^4\BQ+6(\pl^2\BQ)^2-8t_2\pl^2\BQ+3\pl^2_2\BQ 
 +4\pl_w\pl\BQ=0,&
 \end{flalign}
\begin{flalign}
&(\pl_2\pl^2-2t_2\pl_2-3(\vr-w\pl_w)+1)\pl\BQ+6(\pl^2\BQ)(\pl_2\pl\BQ) -2\pl_2\pl_w\BQ=0,&
\end{flalign}
\begin{flalign}
& (\vr-w\pl_w -2t_2\pl_2-2)\pl^2\BQ+\pl_2^3\BQ+2\{\pl_2\pl\BQ,\pl^2\BQ\}_{\pl}
+2\pl_2 \pl_w \pl \BQ=0, &
\end{flalign}
\begin{flalign}
&\pl^4 U+3\pl^2_2U+6\pl^2 U^2+4\pl_w\pl   U=0,\qquad U=\pl^2\BQ-\mbox{$\frac{2}{3} $}t_2 &
 \end{flalign}
Notice that for $n=0$, the function $\BQ$ satisfies these equations, but without the terms containing $\pl_w$.
\end{corollary}
Note that Theorem \ref{theorem9} and Corollary \ref{corolequations} immediately imply Theorem \ref{introtheorem}.

\noindent{\it Proof of Theorem \ref{theorem9}:} \, We first give the proof for $n=0$. Notice that 
\be
\frac{1}{2\tau^2}\prod_i\pl_i^{\ell_i}\tau\circ\tau=\left\{  \begin{array}{l}
0,\quad\mbox{~for ~$\displaystyle{\sum_i}\ell_i$ odd.}\\
\displaystyle{\prod_i\pl_i^{\ell_i}\log\tau} +\mbox{other partials of}~ \log \tau,~~
%
\mbox{~for $\displaystyle{\sum_i}\ell_i=$ even}.
\end{array} \right.
\label{82}\ee
Taking into account this remark, one has for the Hirota equations (\ref{75'}):\\
 {(i)} The partials $\pl_p$ can be ignored, since the $\tau$-functions appearing in 
 $$\det(I-K^{(p)}_{x,{\mathbf t}}\raisebox{1mm}{$\chi$}{}_{E} )=\frac{\tau_E(t)}{\tau(t)}$$ are $\tau$-functions for the p--reduced KP-hierarchy; that is the $\tau$-functions do not contain $t_p,~t_{2p},\ldots$ and thus $\pl_p=0$, when acting on those functions.\\
{(ii)} The odd-degree terms in the Hirota operator (\ref{75'}) do not matter, because they vanish as an Hirota symbol acting on $\tau\circ\tau$, by (\ref{82}).\\
{(iii)} The Hirota equations $\BY_{p+1}$ and $\BY_{1,p}$ contain $\pl_{p+2}$ and $\pl_1\pl_{p+1}$ and many other terms involving $\pl_i$ for $1\leq i\leq p-1$. The partial $\pl_{p+2}$ can be omitted, since it is of odd degree, and $\pl_1\pl_{p+1}$  will be taken care of.\\
{(iv)} The Hirota equations $\BY_{p+2},~\BY_{1,p+1}$ contain $\pl_{p+3},\pl_1^2\pl_{p+1}$, $\pl_1\pl_{p+2}$ and $\pl_2\pl_{p+1}$. The first two terms do not matter, since they are of odd degree, and the exact linear combination $2(p+1)\BY_{p+2}+p\BY_{1,p+1}$ removes the term $\pl_1\pl_{p+2}$, but $\pl_2\pl_{p+1}$ will need to be taken care of.\\
To conclude, the list of Hirota equations (\ref{75'}), acting on p--reduced $\tau$-functions only involves $\pl_1,\ldots, \pl_{p-1}$ and derivatives with regard to higher $t_i$'s only through $\pl_1\pl_{p+1}$ and $\pl_2\pl_{p+1}$.\\
From the Proposition \ref{prop4}, in particular (\ref{Virasoro}) for $n=0$, both $g:=\log\tau_E({\mathbf t}),\log\tau({\mathbf t})$ satisfy
\be
\pl g=\left(\frac{1}{p}\sum_{i\geq p+1}it_i~\pl_{i-p}+\pl_1\right)g+\Gamma_p=:\dt g+\Gamma_p
\label{83}\nonumber\ee
\be
\vr g=\left(\frac{1}{p}\sum_{i\geq1}it_i~\pl_{i}+\pl_{p+1}\right)g+c_p=:\hat\dt g+c_p,
\label{84}\nonumber\ee
where (refering to (\ref{Virasoro}))
%
%
$$
\Gamma_p:=\frac{1}{2p}\sum_{i+j=p}(it_i)(jt_j)-\frac{p-1}{2}t_p,\quad c_p=c_{p,1}^{(0)}=\frac{p^2-1}{12p^2}
$$
and of course $\pl\ln\tau=0$, $\vr\ln\tau=0$. Using $[\pl_i,\dt]=0$ for $2\leq i\leq p-1$, compute inductively
\be
\pl^{i+1}g=\pl^i\pl g=\pl^i(\dt g+\Gamma_p)=\dt\pl^ig=\dt(\dt^ig+\dt^{i-1}\Gamma_p)=\dt^{i+1}g+\dt^i\Gamma_p,
\label{85}\nonumber\ee
\be
\prod_{i=2}^{p-1}\pl_i^{\ell_i}\pl^{\ell_1}g=\prod_{i=2}^{p-1}\pl_i^{\ell_i}(\dt^{\ell_1}g+\dt^{\ell_1-1}\Gamma_p)=\dt^{\ell_1}\prod_{i=2}^{p-1}\pl_i^{\ell_i}g+\dt^{\ell_1-1}\prod_{i=1}^{p-1}\pl_i^{\ell_i}\Gamma_p,
\label{86}\nonumber\ee
\be
\vr\pl g=\vr(\dt g+\Gamma_p)=\dt\vr g=\dt(\hat\dt g+c_p)=\dt\hat\dt g,
\label{87}\nonumber\ee
\be
\pl_2\vr g=\pl_2(\hat\dt g+c_p)=\pl_2\hat\dt g,
\label{87'}\nonumber\ee
and thus on the locus $\LR :=\{t_1=0,t_p=t_{p+1}=... =0\}$,
\be
\dt\hat\dt\big\vert_{\LR}=\pl_1\pl_{p+1}+\frac{1}{p}\left(\pl_1+\sum^{p-1}_{i=2}it_i\pl_i\pl_1\right),\pl_1g=\pl g-\Gamma_p\big\vert_{\LR}.
\nonumber\ee
So we conclude from the formulas above that, on $\LR$:
\bea
 \prod_1^{p-1}\pl_i^{\ell_i}g&=&\prod_{i=2}^{p-1}\pl_i^{\ell_i}\pl ^{\ell_1}g-\frac{1}{2p}\left(\pl_1^{\ell_1-1}
\prod_2^{p-1} \pl_i^{\ell_i}\sum_{i+j=p}(it_i)(jt_j)\right)\Big\vert_{t_1=0},
\label{88a}\\
\nonumber\\
 p\,\pl_1\pl_{p+1}g&=&(p\,\vr -1)\pl g-\sum_2^{p-1}it_i\pl_i\pl_1g+\frac{1}{2p}\sum_{i+j=p\atop{i,j>1}}(it_i)(jt_j)\nonumber\\
 &=&(p\,\vr -1)\pl g-\sum_2^{p-1}it_i(\pl_i\pl g-\frac{i}{p}(p-i)t_{p-i})+\frac{1}{2p}\sum_{i+j=p\atop{i,j>1}}(it_i)(jt_j),\nonumber \\\label{88b}\\
\nonumber\\
 \pl_2\pl_{p+1}g&=&\pl_2\vr g-\frac{1}{p}\left(2\pl_2+\sum_2^{p-1}it_i\pl_i\pl_2\right)g.\label{88f}
 \\
 \nonumber\\
\pl_1g&=&\pl g-\frac{1}{2p}\sum_{i+j=p\atop{i,j>1}}(it_i)(jt_j),\\
\pl_1^2g&=&\pl ^2g-(1-\dt_{2,p})\frac{p-1}{p}t_{p-1}\label{88c},\\
\nonumber\\
\pl^3_1g&=&\pl^3 g-\frac{\dt_{2,p}}{2},\qquad\pl^i_1g=\pl^ig,\quad i\geq 4\label{88d}.\\
\nonumber
%
\eea
Substituting (\ref{88a})-(\ref{88d}) in the explicit PDEs $\BY_{\ell}$ or $\BY_{1,\ell -1}$ in $g=\ln\tau_E(t)$ yields an explicit PDE in $\pl_2,\pl_3,\ldots,\pl_{p-1},\pl ,\vr$, for both $\log\tau_E(t)$ and $\tau(t)$; since $\BQ=\log \tau_E-\log \tau$, one finds two PDEs:
$$\Gamma_0 (\log\tau(t))=0,~~~\Gamma_E(\log\tau_E(t))=\Gamma_E(\BQ+\log \tau(t))=0.
$$
Then form the PDE
$$\Gamma_E(\log\tau_E(t))-\Gamma_0 (\log\tau(t))=
\Gamma_E(\BQ+\log \tau(t))-\Gamma_0 (\log\tau(t))=0
$$
in which $\displaystyle{\prod_2^{p-1}\pl_i^{\ell_i}\log\tau(t)}$ remains; then use (\ref{72}) to explicitly substitute its value as a polynomial in $t_2,...,t_{p-1}$. Carrying out the program in a few cases yields the equations in the corollary \ref{corolequations}, case $n = 0$.\\
For $n > 0$, the only change in the Virasoro, in comparing (\ref{Virasoro}) for $n=0$ and $n >0$, is that $\Gamma_p\mapsto\Gamma_p +nt_p+c_{p,0}^{(1)}$, and  $\vr\rg\vr -w\frac{\pl}{\pl w}=\vr'$ and $c_{p,j}^{(1)}$ will have a different value, and these changes have no actual effect beyond changing $\vr$ to $\vr'$; then carrying out the program will add some extra terms to the equations. This proves the PDEs for $E\subset \BR^+,$ and then we extend the PDEs to $E\subseteq \BR$ by analytic continuation.\qed
In the following appendix we go into more explicit detail and work out a typical example.

\appendix  
\section{Elaboration of the proof of the Theorem \ref{theorem9}}
  
In this appendix, in Lemma \ref{A2}, we prove (\ref{76}); in Lemma \ref{A1} we give the additional formulas beyond (\ref{88a})--(\ref{88d}) necessary in proving the full Corollary \ref{corolequations}.
  
  \begin{lemma}\label{A1} The Hirota symbols corresponding to the coefficients of Lemma \ref{lemma6}, with the noncontributing odd terms removed are, up to a constant, as follows
\be
\begin{array}{lll} 
\BY_3& :&-4\pl_1\pl_3+3\pl_2^2+\pl_1^4\\
\\
\BY_4&: &-3\pl_1\pl_4+2\pl_2\pl_3+\pl_2\pl_1^3\\
\\
\BY_5&: &\frac{1}{4}\pl_2\pl_4-\frac{3}{5}\pl_1\pl_5+\frac{1}{9}\pl^2_3+\frac{1}{9}\pl_1^3\pl_3+\frac{1}{8}\pl_1^2\pl^2_2+\frac{1}{360}\pl_1^6\\
\\
\BY_{1,4}& :&-\frac{1}{8}\pl_2\pl_4+\frac{1}{10}\pl_1\pl_5+\frac{1}{18}\pl_3^2-\frac{1}{36}\pl_1^3\pl_3-\frac{1}{360}\pl^6_1\\
\\
4\BY_{1,4}+10\BY_5& :&\frac{1}{2}\pl_2\pl_4-2\pl_1\pl_5+\frac{2}{3}\pl_3^2+\frac{1}{3}\pl^3_1\pl_3+\frac{1}{2}\pl_1^2\pl^2_2
\end{array}
\label{69}\ee
whose action on $\tau\circ\tau$ yields the following differential equations for $U=\log\tau$.

\be
\begin{array}{lll} 
\BY_3& :&\pl^4_1U+6(\pl_1^2U)^2+3\pl^2_2U-4\pl_1\pl_3U=0\\
\\
\BY_4&: &-3\pl_1\pl_4U+2\pl_2\pl_3U+\pl_1^3\pl_2U+6(\pl_1^2U)(\pl_1\pl_2U)=0\\
\\
\BY_5&: &-\dis\frac{108}{5}\pl_1\pl_5U+\dis\frac{1}{10}\pl_1^6U+6(\pl^2_1U)^3+3(\pl_1^4U)(\pl^2_1U)\\
\\
& &+~9\pl_2\pl_4U+4\pl^2_3U+4\pl_1^3\pl_3U+24(\pl_1^2U)(\pl_1\pl_3 U)\\
\\
& &+~9(\pl_1^2U)(\pl_2^2U)+\frac{9}{2}\pl_1^2\pl_2^2U+18(\pl_1\pl_2U)^2 = 0\\
\\
\BY_{1,4}& :&-\dis\frac{36}{5}\pl_1\pl_5U+\dis\frac{1}{5}\pl^6_1U+12(\pl_1^2U)^3+6(\pl^4_1U)(\pl^2_1U)+9\pl_2\pl_4U\\
\\
& &-~4\pl^2_3U+2\pl^3_1\pl_3U+12(\pl^2_1U)(\pl_1\pl_3U)=0\\
\\
4\BY_{1,4}+10\BY_5& :&-4\pl_1\pl_5U+\pl_2\pl_4U+\frac{4}{3}\pl_3^2U+\dis\frac{2}{3}\pl_1^3\pl_3U+4(\pl_1^2U)(\pl_1\pl_3U)\\
\\
& &+~\pl^2_1\pl_2^2U+4(\pl_1\pl_2U)^2+2(\pl_1^2U)(\pl_2^2U)=0.
\end{array}
\label{70}\ee
\end{lemma}
\noindent{\it Proof}: Equations \ref{69} follow immediately from the definition of Schur polynomials and (\ref{68}), while (\ref{70}) follows from (\ref{69}) and the definition of the Hirota symbol. \qed

As a typical example of explicitly carrying out of the program given in the proof of Theorem \ref{theorem9} we prove (\ref{76}) of Corollary \ref{corolequations}.

\begin{lemma}\label{A2}For all $p\geq 2$, the $\BY_3$-KP  hierarchy member yields the following equation for $\BQ$:

\bean
\lefteqn{\pl^4\BQ+6(\pl^2\BQ)^2+\dt_{2,p}(2-4\vr)\pl\BQ}\\
\\
&+(1-\dt_{2,p})(3\pl^2_2\BQ-(12(\frac{p-1}{p})t_{p-1}\pl+4(1-\dt_{3,p})\pl_3)\pl\BQ)=0.
\eean
\end{lemma}

\noindent{\it Proof of Lemma \ref{A2}:} \, From (\ref{70}) and (\ref{88a})-(\ref{88d}) conclude that $g=\ln\tau_E(\bar{\mathbf t})$ and $\ln\tau(\bar{\mathbf t})$ (remember $\tau(\bar{\mathbf t}) = \tau_0^p(\bar{\mathbf t})$ of Lemma \ref{lemma8}) satisfy

\bean
\BY_3&:&\pl_1^4g+6(\pl_1^2g)^2+3\pl_2^2g-4\pl_1\pl_3g=0,
\\
\pl_1^2g&=&\pl^2g-(1-\dt_{2,p})\left(\frac{p-1}{p}\right)t_{p-1},\qquad\pl_1^4g=\pl^4g,\\
\\
\pl_1\pl_3g&=&\frac{1}{2}(2\vr-1)\pl g,\qquad p=2,\\
\\
\pl_1\pl_3g&=&\left(\pl_3\pl g-\frac{3}{p}(p-3)t_{p-3}(1-\dt_{4,p})\right)(1-\dt_{3,p}),\qquad p>2\\
\\
\pl_2^2g&=&\pl_2^2g(1-\dt_{2,p}),
\eean
and substituting the last 4 relations into the first yields

\bean
 \Gamma_3(g)&:=&\pl^4 g+6\left(\pl^2 g-\left(\frac{p-1}{p}\right)t_{p-1}(1-\dt_{2p})\right)^2+3(1-\dt_{2p})\pl^2_2g\\
& &-4(1-\dt_{3p})\left[\dt_{2,p}\left(\vr -\frac{1}{2}\right)\pl g+(1-\dt_{2,p})(\pl_3\pl g-\frac{3}{p}(p-3)t_{p-3})(1-\dt_{4,p})\right]=0.
\eean
Set $g=g_0=\ln\tau$ and then, since $\pl g_0=\vr g_0=0$, conclude that

\bean
\Gamma_3(g_0)&=&6(1-\dt_{2,p})\left(\left(\frac{p-1}{p}\right)t_{p-1}\right)^2+3(1-\dt_{2,p})\pl_2^2g_0\\
& &\hspace*{4cm}+12(1-\dt_{3,p})(1-\dt_{2,p})\left(\frac{p-3}{p}\right)t_{p-3}=0.
\eean
Since $g=\ln\tau_E(t)=\BQ+\ln\tau_0^{(p)}(t):=\BQ+g_0$, and $\pl g_0=\vr g_0=0$, conclude from the above that

\bean
0&=&\Gamma_3(\BQ+g_0)-\Gamma_3(g_0)\\
\\
&=&\pl^4 \BQ+6(\pl^2 \BQ)^2-12(1-\dt_{2,p}) \frac{(p-1)}{p}t_{p-1}\pl^2 \BQ\\
\\
& &\hspace{4cm}+~6 \left(\frac{(p-1)}{p}t_{p-1}\right)^2(1-\dt_{2,p})\\
\\
& &\hspace{4cm}+~3(1-\dt_{2,p})\pl_2^2\BQ+3(1-\dt_{2,p})\pl^2_2g_0\\
\\
& &\hspace{4cm}-~4(1-\dt_{3,p})\left[\dt_{2p}(\vr -\frac{1}{2})\pl g+   (1-\dt_{3,p})\pl_3\pl \BQ\right]\\
\\
& &\hspace{4cm}+~12(1-\dt_{3,p})(1-\dt_{2,p})  \frac{(p-3)}{p} t_{p-3}-\Gamma_3(g_0)\\
\\
&=&\pl^4\BQ+6(\pl^2\BQ)^2+\left(-12\frac{(p-1)}{p}t_{p-1}\pl^2\BQ+3\pl^2_2\BQ\right)(1-\dt_{2,p})\\
\\
& &\hspace*{2cm}-~4(1-\dt_{3,p})\left[ \dt_{2,p}\left(\vr -\frac{1}{2}\right)\pl g+(1-\dt_{2p})\pl_3\pl \BQ\right],
\eean
which proves Lemma \ref{A2}.\qed
The rest of corollary \ref{corolequations} is proven in the same fashion using Lemma \ref{A1} and (\ref{88a})--(\ref{88d}).

\bibliographystyle{plain}
\bibliography{/Users/mattiacafasso/Documents/BibDeskLibrary.bib}

\end{document}

%% file: SAVE.pdf_t
\begin{picture}(0,0)%
\includegraphics{SAVE.pdf}%
\end{picture}%
\setlength{\unitlength}{3947sp}%
\begingroup\makeatletter\ifx\SetFigFont\undefined%
\gdef\SetFigFont#1#2#3#4#5{%
  \reset@font\fontsize{#1}{#2pt}%
  \fontfamily{#3}\fontseries{#4}\fontshape{#5}%
  \selectfont}%
\fi\endgroup%
\begin{picture}(17469,5274)(-5606,-6373)
\put(-3884,-3556){\makebox(0,0)[lb]{\smash{{\SetFigFont{12}{14.4}{\rmdefault}{\mddefault}{\updefault}{\color[rgb]{0,0,0}$\pi/3$}%
}}}}
\put(3001,-3511){\makebox(0,0)[lb]{\smash{{\SetFigFont{12}{14.4}{\rmdefault}{\mddefault}{\updefault}{\color[rgb]{0,0,0}$\pi/4$}%
}}}}
\put(1351,-3511){\makebox(0,0)[lb]{\smash{{\SetFigFont{12}{14.4}{\rmdefault}{\mddefault}{\updefault}{\color[rgb]{0,0,0}$\pi/4$}%
}}}}
\put(9901,-3661){\makebox(0,0)[lb]{\smash{{\SetFigFont{12}{14.4}{\rmdefault}{\mddefault}{\updefault}{\color[rgb]{0,0,0}$\pi/5$}%
}}}}
\put(8101,-3661){\makebox(0,0)[lb]{\smash{{\SetFigFont{12}{14.4}{\rmdefault}{\mddefault}{\updefault}{\color[rgb]{0,0,0}$2\pi/5$}%
}}}}
\put(2551,-1411){\makebox(0,0)[lb]{\smash{{\SetFigFont{12}{14.4}{\rmdefault}{\mddefault}{\updefault}{\color[rgb]{1,0,0}$\Gamma_3^+=i\BR$}%
}}}}
\put(4201,-2311){\makebox(0,0)[lb]{\smash{{\SetFigFont{12}{14.4}{\rmdefault}{\mddefault}{\updefault}{\color[rgb]{0,0,1}$\Gamma_3^-$}%
}}}}
\put(4051,-5386){\makebox(0,0)[lb]{\smash{{\SetFigFont{12}{14.4}{\rmdefault}{\mddefault}{\updefault}{\color[rgb]{0,0,1}$\Gamma_3^-$}%
}}}}
\put(10351,-1861){\makebox(0,0)[lb]{\smash{{\SetFigFont{12}{14.4}{\rmdefault}{\mddefault}{\updefault}{\color[rgb]{1,0,0}$\Gamma_4^+$}%
}}}}
\put(10876,-2986){\makebox(0,0)[lb]{\smash{{\SetFigFont{12}{14.4}{\rmdefault}{\mddefault}{\updefault}{\color[rgb]{0,0,1}$\Gamma_4^-$}%
}}}}
\put(8251,-1861){\makebox(0,0)[lb]{\smash{{\SetFigFont{12}{14.4}{\rmdefault}{\mddefault}{\updefault}{\color[rgb]{0,0,1}$\Gamma_4^-$}%
}}}}
\put(7051,-2536){\makebox(0,0)[lb]{\smash{{\SetFigFont{12}{14.4}{\rmdefault}{\mddefault}{\updefault}{\color[rgb]{1,0,0}$\Gamma_4^+$}%
}}}}
\put(-2324,-2326){\makebox(0,0)[lb]{\smash{{\SetFigFont{12}{14.4}{\rmdefault}{\mddefault}{\updefault}{\color[rgb]{0,0,1}$\Gamma_2^-$}%
}}}}
\put(-2819,-3541){\makebox(0,0)[lb]{\smash{{\SetFigFont{12}{14.4}{\rmdefault}{\mddefault}{\updefault}{\color[rgb]{0,0,0}$\pi/3$}%
}}}}
\put(-4289,-2341){\makebox(0,0)[lb]{\smash{{\SetFigFont{12}{14.4}{\rmdefault}{\mddefault}{\updefault}{\color[rgb]{1,0,0}$\Gamma_2^+$}%
}}}}
\end{picture}%